\documentclass[sigconf]{acmart} 

\usepackage{graphicx}
\usepackage{subfigure}
\usepackage{multirow}
\usepackage[ruled,vlined,linesnumbered]{algorithm2e}

\copyrightyear{2022}
\acmYear{2022}
\setcopyright{acmcopyright}
\acmConference[WWW '22]{Proceedings of the ACM Web Conference 2022}{April 25--29, 2022}{Virtual Event, Lyon, France}
\acmBooktitle{Proceedings of the ACM Web Conference 2022 (WWW '22), April 25--29, 2022, Virtual Event, Lyon, France}
\acmPrice{15.00}
\acmDOI{10.1145/3485447.3511938}
\acmISBN{978-1-4503-9096-5/22/04}

\begin{document}

\title{Trustworthy Knowledge Graph Completion Based on Multi-sourced Noisy Data}

%

\settopmatter{authorsperrow=2}

\author{Jiacheng Huang, Yao Zhao, Wei Hu}
\authornote{Wei Hu is the corresponding author.}
\affiliation{
    \department{State Key Laboratory for Novel Software Technology}
    \department{National Institute of Healthcare Data Science}
    \institution{Nanjing University}
    \city{Nanjing}
    \country{China}
}
\email{{jchuang,yaozhao}.nju@gmail.com,  whu@nju.edu.cn}

\author{Zhen Ning, Qijin Chen, Xiaoxia Qiu}
\author{Chengfu Huo, Weijun Ren}
\affiliation{
    \institution{Alibaba Group}
    \city{Hangzhou}
    \country{China}
}
\email{{ningzhen.nz,qijin.cqj,xiaoxia.qxx}@alibaba-inc.com}
\email{{chengfu.huocf,afei}@alibaba-inc.com}










\renewcommand{\shortauthors}{J. Huang, Y. Zhao, W. Hu, Z. Ning, Q. Chen, X. Qiu, C. Huo, and W. Ren}

\begin{abstract}
Knowledge graphs (KGs) have become a valuable asset for many AI applications. 
Although some KGs contain plenty of facts, they are widely acknowledged as incomplete. 
To address this issue, many KG completion methods are proposed. 
Among them, open KG completion methods leverage the Web to find missing facts. 
However, noisy data collected from diverse sources may damage the completion accuracy. 
In this paper, we propose a new trustworthy method that exploits facts for a KG based on multi-sourced noisy data and existing facts in the KG. 
Specifically, we introduce a graph neural network with a holistic scoring function to judge the plausibility of facts with various value types. 
We design value alignment networks to resolve the heterogeneity between values and map them to entities even outside the KG. 
Furthermore, we present a truth inference model that incorporates data source qualities into the fact scoring function, and design a semi-supervised learning way to infer the truths from heterogeneous values. 
We conduct extensive experiments to compare our method with the state-of-the-arts. 
The results show that our method achieves superior accuracy not only in completing missing facts but also in discovering new facts.
\end{abstract}

\ccsdesc[500]{Information systems~Data extraction and integration}
\ccsdesc[300]{Computing methodologies~Neural networks}

\keywords{Knowledge graph completion; truth inference; noisy data}

\maketitle

\section{Introduction}
Knowledge graphs (KGs), which represent real-world facts in the form of triples, have many important applications, e.g., semantic search, question answering and recommender systems. 
Some KGs, e.g., DBpedia~\cite{DBpedia}, Freebase~\cite{Freebase}, Wikidata~\cite{Wikidata} and Probase~\cite{Probase}, are quite large; however, they are still acknowledged as incomplete \cite{TransE}.
Existing studies cope with a task called \emph{KG completion} to discover missing facts for a KG (see surveys~\cite{KGESurvey,KGSurvey,LP1}).
The majority of recent works focus on learning the embeddings of entities and relations in the KG, and leveraging them to predict the relations between two entities (or predict an entity given another entity and a relation). 
However, it is arguable to assume that all the entities and relations have been covered by the KG, as most KGs are built based on the open-world assumption \cite{AI}.
In fact, there are many other entities outside the KG but on the Web, and a few \emph{open KG completion} methods~\cite{OKELE,KnowledgeVault,MIA,OWE,ConMask} attempt to mine facts from web pages.
To validate the plausibility of mined facts, these methods are challenged by multi-sourced noisy data with varying quality. 

\smallskip
\noindent\textbf{Current approaches.}
Conventional KG completion methods, e.g., \cite{TransE,BoxE,DualE,ConvEX,M2KGNN}, learn embeddings for entities and relations within a KG, and define scoring functions to measure the plausibility of facts based on the embeddings. 
Most methods focus on completing relational facts between entities, e.g., $(\textit{Boris Johnson},\textit{nationality},\textit{UK})$, whereas facts in real-world KGs contain other types of values, e.g., number, datetime, category and string. 
As far as we know, only a few methods \cite{MTKGNN,KBLRN} can process one or two extra types of values, but they are limited to completing the values inside the KG.

Open KG completion methods~\cite{OKELE,KnowledgeVault,OWE,ConMask,MIA} can mine new facts containing unseen entities, which appear in web pages but may not exist in the KG. 
In this way, open KG completion methods broaden the extent of KG by leveraging external resources such as online encyclopedias and vertical websites. 
The main challenge is the noises from multi-sourced data, which severely interfere the quality of discovered facts. 
Several existing methods~\cite{KnowledgeVault,OWE,ConMask} focus on adding new facts that do not conflict with existing ones in the KG. 
Their shortcoming is that they ignore the quality of data sources.
For example, data on Wikipedia are carefully validated, but data on personal websites may be unreliable. 
A recent solution, OKELE~\cite{OKELE}, models the data source quality through a probabilistic graphical model. However, it cannot leverage prior knowledge in the KG to judge the plausibility of new facts. 

\begin{figure}
    \centering
    \includegraphics[width=\columnwidth]{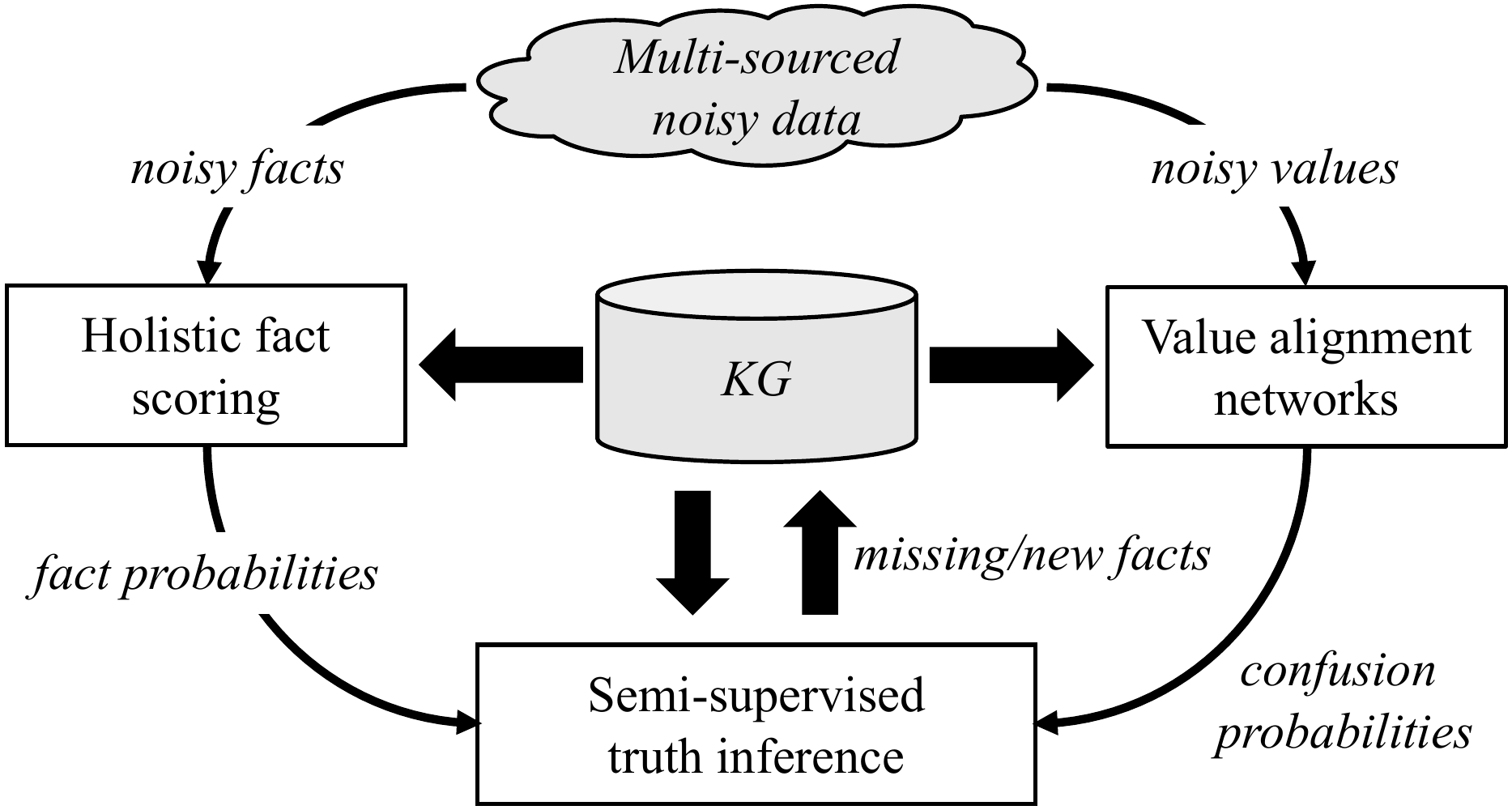}
    \caption{Workflow of the method}
    \label{fig:workflow}
\end{figure}

\smallskip
\noindent\textbf{Our approach.} In this paper, we propose a novel trustworthy KG completion method called \textbf{TKGC}. 
Our fundamental idea is to leverage noisy data from diverse web pages and prior knowledge in a KG symbiotically, so that KG completion can benefit from the open Web to not only add missing facts inside the KG, but also discover new facts outside the KG; meanwhile data noises can be resolved by considering existing facts in the KG. 
Figure~\ref{fig:workflow} depicts the workflow of our method, which consists of three main components to address key challenges:

\emph{Holistic fact scoring} measures the plausibility of facts extracted from multi-sourced noisy data by learning embeddings to represent entities, attributes and values. 
In addition to relational facts, we propose a graph neural network (GNN) with a holistic fact scoring function to handle facts with various types of values. 

\emph{Value alignment networks} softly resolve the heterogeneity between values, e.g., ``UK'' versus ``United Kindom'', by a literal-literal alignment network. 
Furthermore, literals are mapped into the entity embedding space by a literal-entity alignment network, so that facts with unseen entities can be discovered. 

\emph{Semi-supervised truth inference} identifies which noisy facts are plausible and adds the trustworthy ones into the KG. 
It defines the confusion probabilities to capture the relevance between aligned values and infers the truths by a semi-supervised learning process with the holistic fact scoring. 

In summary, our main contributions in this paper are fourfold:
\begin{itemize}
  \item We introduce a holistic fact scoring model to measure the plausibility of facts with different types of values such as entity, number and string. (Section~\ref{sec:scoring})
  
  \item We design value alignment networks to resolve heterogeneous values from multiple sources, and align values with entity embeddings to discover facts with unseen entities. (Section~\ref{sec:align})
  
  \item We incorporate a new truth inference model into the fact scoring model to realize trustworthy KG completion. 
  Furthermore, we design a semi-supervised learning method to infer plausible facts from noisy data with a KG providing prior knowledge. (Section~\ref{sec:infer})
  
  \item We carry out extensive experiments to evaluate the effectiveness of our method TKGC. 
  Our experimental results show that TKGC achieves superior performance on a benchmark dataset. 
  Compared with the best-performing competitor, it particularly raises 0.048 of F1-score on relational fact completion and reduces 0.020 of RMSE (root of mean square error) on literal fact completion. (Section~\ref{sec:exp})
\end{itemize}


\section{Preliminaries}
\label{sec:prelim}

\subsection{Problem Statement}
A KG collects a wealth of structured facts. It uses entities to denote real-world objects and describes the facts about entities in the form of $(entity, attribute, value)$, or $(e,a,v)$ for short. 
The value in each fact can be either an entity or a literal. 
A fact with an entity as the value is called a \emph{relational fact}, and a fact with a literal as the value is called a \emph{literal fact}.
Therefore, a KG $K$ can be defined as a 5-tuple $(E, A, L, T_{rel}, T_{lit})$, where $E$ is the entity set, $A$ is the attribute set, $L$ is the literal set, $T_{rel}\subseteq E\times A\times E$ is the relational fact set, and $T_{lit}\subseteq E\times A\times L$ is the literal fact set. We denote the value set by $V=L\cup E$.
For each attribute $a$, we denote the domain of its possible values by $V_a=\{v\in V\,|\,\exists e\in E,(e,a,v)\in T_{rel}\cup T_{lit}\}$. 
For example, given $a=\textit{gender}$, we have $V_a = \{\text{``male''}, \text{``female''}\}$. 

A data source $s$ provides a collection of noisy facts called \emph{claims} that may be inaccurate or incorrect. 
A claim extracted from a data source $s$ is denoted by $(e, a, v, s)$, which means that source $s$ claims ``attribute $a$ of entity $e$ has value $v$''. 
Trustworthy KG completion aims to resolve inconsistent claims (e.g., different values of a given pair of entity and attribute) from multiple data sources, and identify the trustworthy claims called \emph{truths}. 

\smallskip
\noindent\textbf{Problem definition.} Given a KG $K=(E, A, L, T_{rel}, T_{lit})$ as prior knowledge and a set of claims $C=\big\{(e,a,v,s)_i\big\}_{i=1}^{I}$ extracted from a set of data sources $S=\{s_j\}_{j=1}^{J}$, the task of trustworthy KG completion is to identify the truths $C'\subseteq C$ and add them into $K$.

\subsection{Related Work}

\noindent\textbf{Conventional KG completion.} Recent KG completion methods mainly use the embedding techniques to complete missing relational facts~\cite{LP1,KGESurvey,KGSurvey}. 
There are three general types of methods: geometric models~\cite{TransE,BoxE,DualE}, tensor decomposition models \cite{DistMult,TuckER} and deep learning models~\cite{ConvE,CapsE,RSN,RGCN,SACN,CompGCN,M2KGNN,ConvEX}. 
Inspired by the recent achievement of GNNs in KG completion (e.g., \cite{RGCN,SACN,CompGCN,M2KGNN}), we choose a GNN model for fact scoring. 
Additionally, a few KG completion methods, e.g., DKRL~\cite{DKRL} and LiteralE~\cite{LiteralE}, encode literals to vectors and add them into entity embeddings.
However, they are not designed for completing literal facts, rather using literals to improve the quality of relational fact completion. 
MT-KGNN~\cite{MTKGNN} and KBLRN~\cite{KBLRN} both leverage different fact scoring functions for entities and numbers. 
MT-KGNN uses multi-task learning, while KBLRN proposes a ``product of experts'' approach.
Generally, conventional KG completion methods cannot complete facts with unseen entities~\cite{OWE,ConMask}.

\smallskip
\noindent\textbf{Open KG completion.} By leveraging external knowledge, open KG completion methods~\cite{OWE,ConMask,KnowledgeVault,OKELE} can find new facts with unseen entities. ConMask~\cite{ConMask} fuses entity names and descriptions to obtain entity embeddings, and combines entity embeddings and relation embeddings to judge the plausibility of relational facts. 
OWE~\cite{OWE} extends the conventional KG completion methods, and trains a neural network to map the name embeddings of unseen entities to the KG embedding space. 
However, ConMask and OWE do not consider the noises derived from external data sources. 
To address the noise problem, Knowledge Vault~\cite{KnowledgeVault} validates new facts by mining rules with a path ranking algorithm and matrix completion with tensor decomposition. 
OKELE~\cite{OKELE} first predicts missing attributes for long-tail entities, and then extracts true facts with a probabilistic graphical model. 
Knowledge Vault focuses on leveraging prior knowledge in the KG, while OKELE aims to reduce noises of external knowledge from the Web. 
Both of them may suffer from error accumulation due to their pipeline architecture.
Compared with them, TKGC makes use of both prior and external knowledge in a semi-supervised manner.

\smallskip
\noindent\textbf{Truth inference.} Truth inference methods are proposed to infer truths from multi-sourced noisy data~\cite{TISurvery,TDSurvey}.
Majority voting~\cite{TISurvery} is a very simple approach choosing the majority as the truth. 
It treats all data sources equally, therefore many untrustworthy claims would generate inaccurate truths. 
To overcome this issue, previous studies~\cite{TruthFinder,PooledInvestment,CATD} leverage the quality of a data source to represent the ability that the data source provides correct claims, and infer truths based on the high-quality data sources. 
LTM~\cite{LTM}, LCA~\cite{LCA}, MBM~\cite{MBM} and BWA~\cite{BWA} further use probabilistic models to measure the trustworthiness of each claim. 
Compared with all of them, TKGC can leverage prior knowledge provided in the KG to improve the accuracy of truth inference. 
For instance, \textit{nationality} and \textit{birth place} of the same person are highly relevant. 
Currently, only TKGC can capture this fact relevance.
Furthermore, TKGC can make subtle comparisons using the value alignment networks to capture value relevance.

\section{Holistic Fact Scoring}
\label{sec:scoring}

Inspired by embedding-based KG completion methods~\cite{KGESurvey,KGSurvey,LP1}, we encode the knowledge in a KG through representing its entities, attributes and values as embeddings and learning scoring functions over them to judge whether a fact is plausible.
However, the majority of existing methods, such as \cite{TransE,BoxE,DualE,DistMult,TuckER,ConvE,CapsE,RSN,RGCN,SACN,CompGCN,M2KGNN,ConvEX}, only address relational fact completion. 
Unlike them, we propose a holistic fact scoring function $\mathcal{F}:T_{rel}\cup T_{lit}\mapsto [0,+\infty)$ to model both relational triples and literal triples, such that $\mathcal{F}(e, a, v)\to 0$ if and only if fact $(e, a, v)$ is plausible. 

\subsection{Entity Encoding}

It is widely observed that neighboring entities, especially within two hops, are often critical to discover the missing attribute values \cite{RGCN,AliNet}.
For example, two albums created by the same singer often have the similar genre.
In this paper, we employ a two-layer GNN to model this graph structure. 
Considering the scale of real-world KGs which may contain millions of entities, we choose the GraphSage framework~\cite{GraphSage}, which defines a neighbor sampling function $\mathcal{N}:E\rightarrow 2^E$ to improve the scalability, and an aggregation function $\mathcal{A}:E\times 2^E\rightarrow \mathbb{R}^d$ to aggregate the neighboring information. 
For each entity $e\in E$, its hidden representation at the $k^\text{th}$ layer is denoted by $\mathbf{h}_e^k$. 
At the input layer ($k = 0$), the hidden representation $\mathbf{h}_e^0$ is called \emph{initial entity representation} and denoted by $\mathbf{e}_{init}$. 
The aggregation function $\mathcal{A}()$ transforms $\mathbf{h}_e^{k-1}$ to $\mathbf{h}_e^k$ based on the sampled neighbors $\mathcal{N}(e)$. 
Specifically, we use the mean-based aggregation function, which is defined as
\begin{align}
\begin{split}
  \mathbf{h}_e^k &= \mathcal{A}\big(e, \mathcal{N}(e)\big) \\
                 &= \sigma\Big(\mathbf{W}^k_g \operatorname{MEAN}\big(\{\mathbf{h}_e^{k-1}\}\cup\{\mathbf{h}_{e'}^{k-1}\,|\, e'\in \mathcal{N}(e)\}\big)\Big),
\end{split}
\end{align}
where $\sigma()$ is the sigmoid function, $\mathbf{W}^k_g$ is the weight matrix at the $k^\text{th}$ layer, and $\operatorname{MEAN}()$ is the element-wise mean function.

As the multiple layers of aggregation function make it difficult to learn the initial entity representation $\mathbf{e}_{init}$, we use a residual connection and concatenate the initial representation with the output of residual connection as the final representation:
\begin{align}
\mathbf{e} & = \Big[\mathbf{e}_{init};\mathbf{e}_{init} + \sigma\big(\mathbf{W}_{res}(\mathbf{e}_{init}+\mathbf{h}_e^K)\big)\Big],\label{eq:res}
\end{align}
where $K$ is the number of aggregation layers, $\mathbf{W}_{res}$ is the weight matrix. Dropout and layer normalization are adopted in Eq.~(\ref{eq:res}) to avoid overfitting.

\subsection{Relational Fact Scoring}

We design a multi-class classification model for relational facts and leverage the weighted cross-entropy loss as the scoring function.
A key challenge is the class imbalance problem, and we automatically calculate the value weights to address this problem. 
The scoring function $\mathcal{F}_{rel}()$ for a relational fact $(e, a, v)$ is defined as
\begin{align}\label{eq:score_c}
  \mathcal{F}_{rel}(e, a, v) = -\lambda_{a,v}\cdot \log \frac{\exp(\mathbf{e}^T\mathbf{W}_a\mathbf{v})}{\sum_{v'\in V_a}\exp(\mathbf{e}^T\mathbf{W}_a\mathbf{v'})},
\end{align}
where $\lambda_{a,v}$ is the weight for value $v$ of attribute $a$, $\mathbf{e}$ is the vector representation of entity $e$, $\mathbf{W}_a$ is the matrix representation of attribute $a$, $\mathbf{v}$ and $\mathbf{v'}$ are the vector representations of values $v,v'$, respectively, and $V_a$ is the domain of attribute $a$. 
As $V_a$ may contain thousands of entities, we use a threshold $N_v$ ($N_v < |V_a|$) to restrict the size of $V_a$. 
Specifically, we sample a subset $V'_a$ from $V_a$ which contains $v$ and other $N_v - 1$ elements.

Also, there may be thousands of entity and attribute pairs appearing in the relational fact set. Thus, it is impractical to manually tune weights $\lambda_{a,v}$ for different pairs. 
In this paper, for value $v$ of attribute $a$, $\lambda_{a,v}$ is automatically calculated as follows:
\begin{align}
  \lambda_{a,v} &= \frac{\lambda_{a,v}'}{\sum_{v'\in V'_a}\lambda_{a,v'}'}, \\
  \lambda_{a,v}' &= \frac{1}{\log\big(1 + |\{e\,|\,(e,a,v)\in T_{rel}\}|\big)}.
\end{align}

\subsection{Literal Fact Scoring}
\label{sec:literal}

The general types of values in literal facts include string, category, number and datetime. 
The value type can be obtained or inferred with the corresponding attribute based on the schema or existing facts in the KG.
For a string, we reuse the relational fact scoring function, by replacing the learnable entity embedding with the vector representation of the string encoded with BERT~\cite{BERT}.  
We adopt the pre-trained uncased BERT base model to encode string values.
As generating texts from a KG is still a challenging problem~\cite{TextGeneration}, we select correct string values from claims rather than directly generate them.
For a category, we treat it as an entity. 
For a datetime, we transform it to a real number that denotes the time duration in days from 2000-01-01.
Note that a datetime before 2000-01-01 is transformed into a negative number.
For other value types, we treat them as strings.
Thus, we only need to consider numeric fact scoring. 
The same preprocessing is reused in Section~\ref{sec:infer}. 

For numeric facts, we define the scoring function as the $L_1$-norm loss for a regression model, where attribute embeddings serve as the model parameters. 
Formally, the scoring function $\mathcal{F}_{lit}()$ for a numeric fact $(e, a, v)$ is defined as
\begin{align}\label{eq:score_n}
  \mathcal{F}_{lit}(e, a, v) = \big|\sigma(\mathbf{e}^T\mathbf{a}+b_a) - \operatorname{norm}_a(v)\big|,
\end{align}
where $\sigma()$ is the sigmoid function, $b_a$ is the bias, $\mathbf{e}$ is the vector representation of entity $e$, $\mathbf{a}$ is the vector representation of attribute $a$, and $\operatorname{norm}_a()$ is a normalization function defined as $\operatorname{norm}_a(v)=\frac{v-\min V_a}{\max V_a - \min V_a}$. 
When $\max V_a = \min V_a$, we simply discard the facts containing attribute $a$, as these values are statistically meaningless. 
When $a$ is a $k$-valued attribute, we apply $k$ different embeddings $\{\mathbf{a}_i\}_{i=1}^k$ and biases $\{b_{a,i}\}_{i=1}^k$ to fit $k$ values $\{v_i\}_{i=1}^k$ of entity $e$, and the scoring function is
\begin{align}\label{eq:multi}
    \mathcal{F}_{lit}(e, a, v) = \min_{i=1,2\dots,k} \big|\sigma(\mathbf{e}^T\mathbf{a}_i + b_{a,i}) - \operatorname{norm}_a(v_i)\big|.
\end{align}

We expect to jointly learn entity embeddings based upon both relational and literal facts.
Thus, we use $\mathcal{F}$ to unify $\mathcal{F}_{rel}$ and $\mathcal{F}_{lit}$. 
Given a set of existing facts, we minimize the following fact loss function:
\begin{equation}\label{eq:fact_loss}
  \mathcal{L}_{fact} = \sum_{(e,a,v)\in T_{rel}\cup T_{lit}} \mathcal{F}(e, a, v),
\end{equation}
where $T_{rel}$ is the relational fact set, and $T_{lit}$ is the literal fact set. 

\section{Value Alignment Networks}
\label{sec:align}

As claims are collected from different data sources, values in the claims may be heterogeneous.
In order to produce consistent truths based on facts and claims, we have to find which pairs of extracted values have similar meanings. In addition, as entities in the web pages are represented as texts, we also need to align them.

\subsection{Literal-Literal Alignment}

We first normalize values and then align values with the same type. 
For number and datetime, we conduct the same normalization in Section~\ref{sec:literal}. 
For category, it can also be hand-craftily normalized, since they are enumerable.
However, due to the heterogeneity, it is hard to normalize string, which is our focus. 

Given a pair of strings $v, v'$, we design a literal-literal alignment neural network $\operatorname{LL-ANN}(v, v')$. 
Unlike most existing string alignment methods that are proposed to judge whether two strings are the same, we use $\operatorname{LL-ANN}(v, v')$ to measure whether value $v$ can infer $v'$. 
In this sense, $\operatorname{LL-ANN}(v, v')$ is an asymmetric similarity measure~\cite{Asymmetric}. 
Based on this asymmetric measure, the subsequent truth inference can increase the trustworthiness of strings which can be inferred from other strings. 
Consider a set of extracted values $\{\text{``pop''}, \text{``rock''}, \text{``pop rock''},$ $ \text{``folk rock''}\}$. 
When measuring the accuracy, once the truth inference algorithm decides $\text{``pop rock''}$ or $\text{``folk rock''}$ is correct, $\text{``pop''}$ is also correct. 
This inference can help increase the trustworthiness on $\text{``pop''}$. 

\begin{figure}
    \centering
    \includegraphics[width=\columnwidth]{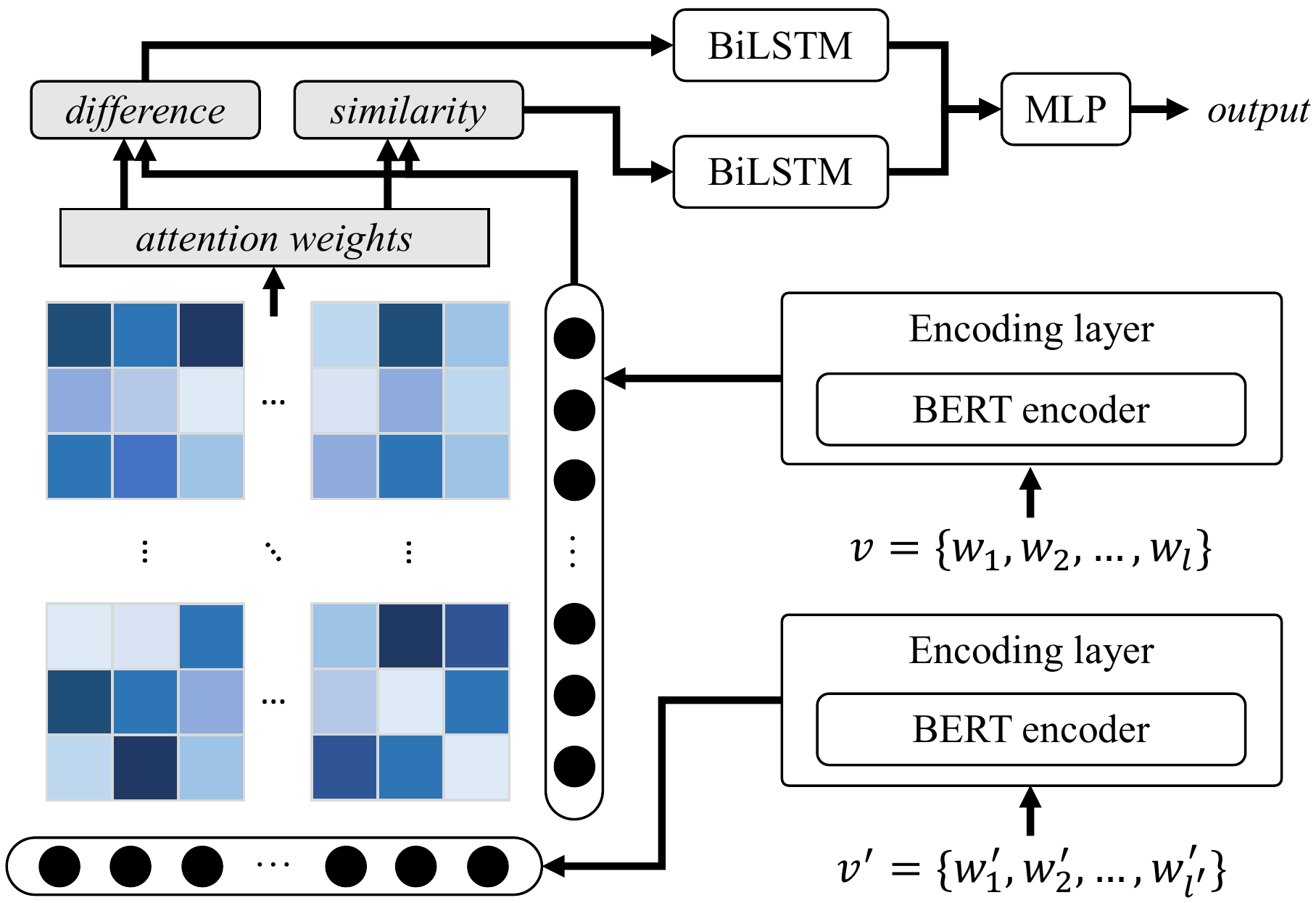}
    \caption{Literal-literal alignment network}
    \label{fig:llann}
\end{figure}

As illustrated in Figure~\ref{fig:llann}, we first encode $v, v'$ into two token embedding sequences, then use an attention mechanism to compare the token embeddings and combine the comparison results with LSTM, and finally use an MLP for classification.

\smallskip
\noindent\textbf{Encoding layer.} In the encoding layer, we follow the typical string matching network \cite{EMTransformer} to encode word sequences with BERT and obtain token embedding sequences. 
We still use the pre-trained uncased BERT base model.
Given a value $v=\{w_i\}_{i=1}^{l}$, where $w_i$ is the $i^\text{th}$ word and $l$ is the total words, the token embedding sequence is $\{\mathbf{w}_1,\mathbf{w}_2,\dots,\mathbf{w}_l\}$. 
For value $v'=\{w'_j\}_{j=1}^{l'}$, the token embedding sequence is defined similarly.

\smallskip
\noindent\textbf{Alignment layer.} In this layer, we align each token embedding $\mathbf{w}_i$ of value $v$ with all token embeddings of value $v'$, and find its aligned token embedding $\mathbf{att}_i$. 
As a result, we obtain the aligned token embedding sequence:
\begin{align}
  \mathbf{att}_i = \sum_{j=1}^{l'} \operatorname{cos}(\mathbf{w}_i,\mathbf{w}'_{j})\mathbf{w}'_{j}\quad (i=1,2,\dots,l), 
\end{align} 
where $\operatorname{cos}()$ is the cosine similarity function for vectors.

Then, to achieve a subtle comparison, we propose to obtain the similarity signals and difference signals with bidirectional LSTM:
\begin{align}
  \mathbf{diff} &= \operatorname{BiLSTM}(\mathbf{w}_1-\mathbf{att}_1, \mathbf{w}_2-\mathbf{att}_2, \dots,\mathbf{w}_l-\mathbf{att}_l), \\
  \mathbf{sim} &= \operatorname{BiLSTM}\Big(\frac{\mathbf{w}_1+\mathbf{att}_1}{2}, \frac{\mathbf{w}_2+\mathbf{att}_2}{2},\dots,\frac{\mathbf{w}_l+\mathbf{att}_l}{2}\Big).
\end{align}

We take the last hidden layer of bidirectional LSTM as output.

\smallskip
\noindent\textbf{Classification layer.} We apply an MLP to combine the similarity and difference signals. 
Finally, the string alignment probability is calculated as follows:
\begin{align}
  \operatorname{LL-ANN}(v, v') = \operatorname{MLP}(\mathbf{sim}, \mathbf{diff}).
\end{align}

\subsection{Literal-Entity Alignment}

A string value may also refer to a real-world object. 
Current entity linking methods~\cite{Falcon,BLINK} can link texts to corresponding entities in the KG, but we also expect to validate the correctness of facts with corresponding entities outside the KG. 
In this paper, we propose a literal-entity alignment neural network $\operatorname{LE-ANN}$ to approximate literal $v$ to its corresponding entity $e$ based on the literal and entity embeddings. In this way, facts about unseen entity $e$, represented by $v$, can be validated by fact scoring based on  $\operatorname{LE-ANN}(v)$.

We use BERT to encode literal $v$ into a vector, and feed the vector into MLP. 
Therefore, $\operatorname{LE-ANN}$ is defined as 
\begin{align}
  \operatorname{LE-ANN}(v) = \operatorname{MLP}\big(\operatorname{BERT}(v)\big),
\end{align}
where $\operatorname{MLP}()$ is a two-layer MLP with sigmoid activation. 
Given an entity embedding $\mathbf{e}$ and a literal value $v$, the literal-entity alignment probability is calculated by $\exp\big(-||\operatorname{LE-ANN}(v) - \mathbf{e}||\big)$.

\section{Semi-supervised Truth Inference}
\label{sec:infer}

While the fact scoring model can complete the KG by maximizing the scoring function $\mathcal{F}(e, a, v)$ for missing value $v$ of attribute $a$ w.r.t. entity $e$, it is restricted to the internal knowledge within the KG. 
To leverage external knowledge from the open Web which contains multi-sourced claims, we propose a truth inference model upon the fact scoring and use a semi-supervised learning method to infer the truths.

Following most truth inference models~\cite{TISurvery,TDSurvey}, we consider two key factors, i.e., the quality of data sources and the trustworthiness of claims. 
We design the confusion probability to model how a data source makes errors. 
Compared with other solutions~\cite{PooledInvestment,CATD,LTM,LCA,MBM,BWA}, the confusion probability makes sophisticated comparison for complex value types. 
We also propose a new observed value probability estimation method to estimate the trustworthiness of claims with prior knowledge provided by the fact scoring function.

\subsection{Confusion Probability} 

Our truth inference model assumes that noises change the truths with an inherent probability distribution determined by a data source. 
The probability that a data source $s$ changes truth value $v^*$ to observed value $v$, called \emph{confusion probability}, is denoted by $\Pr[v\,|\,v^*, s]$. 
We can define a difference function $d(v, v^*)$ to estimate the difference between $v$ and $v^*$. 
We assume that the confusion probability is determined by the difference function, and the difference subjects to a Gaussian distribution with zero mean value and $(k_a\sigma_{s})^2$ variance, i.e., $\Pr[v\,|\,v^*,s]=\Pr[d(v,v^*)]\sim N\big(0,(k_a\sigma_{s})^2\big)$, where $k_a$ is a scale factor and $\sigma_{s}$ reflects the noise caused by data source $s$. 
As the value differences corresponding to different attributes often have different scales, $k_a$ is used to regularize these differences. 
For different types of values, we define different difference functions:
\begin{itemize}
\item For entity (and category), we define the difference function as a metric function in the entity (category) embedding space. 
Given two entities (or categories) $v, v^*$ and their embeddings $\mathbf{v}, \mathbf{v}^*$, we first project them into a latent feature space with dimension $d_{de}$ by a linear transform $\mathbf{W}_e$, and use $L_1$-norm as the distance measure. 
Formally, the difference function is defined as $d(v, v^*) = ||\mathbf{W}_e \mathbf{v} - \mathbf{W}_e \mathbf{v}^*||$. 

\item For number and datetime, we define the difference function as the absolute difference, i.e., $d(v, v^*) = |v - v^*|$. 
With this difference function, our truth inference coincides with the widely-used Gaussian truth inference~\cite{CATD}, which assumes that the observed value subjects to a Gaussian distribution with mean value $v^*$ and variance controlled by data source $s$, i.e., $v\sim N(v^*,k_a\sigma_s)$.

\item For string, we reuse the literal-literal alignment model to obtain the difference $d(v, v^*) = 1 - \operatorname{LL-ANN}(v, v^*)$.
\end{itemize}

\subsection{Observed Value Probability}

Since the truth value $v^*$ is unknown in the truth inference problem, we seek to estimate the conditional probability $\Pr[v\,|\,e,a,s]$ instead of $\Pr[v\,|\,v^*, s]$.
We intend to leverage truth value $v^*$ to bridge the observed value $v$ and the condition $(e, a, s)$. 
Based on the law of total probability, we can integrate over all possible $v^*$ to evaluate the conditional probability:
\begin{align}
    \Pr[v\,|\,e,a,s] = \int_{v^*} \Pr[v\,|\,v^*,e,a,s] \Pr[v^*\,|\,e,a,s] dv^*.
\end{align}

As we assume that the observed value $v$ is directly determined by truth value $v^*$ and data source $s$, $v$ is conditionally independent with $e, a$, i.e., $\Pr[v\,|\,v^*,e,a,s]=\Pr[v\,|\,v^*, s]$. 
Moreover, as the truth value $v^*$ is determined by entity $e$ and attribute $a$, we have $\Pr[v^*\,|\,e,a,s]=\Pr[v^*\,|\,e,a]$. 
Together, we have
\begin{align}
    \Pr[v\,|\,e,a,s] & = \int_{v^*} \Pr[v\,|\,v^*, s] \Pr[v^*\,|\,e,a] dv^*.
\end{align}

While the truth value space is undetermined in truth inference, we can use the possible truth value set $V^*_{e,a}=\{v^*\,|\,(e,a,v^*,s')\in C\}$ to approximate it, where $C$ is the claim set. 
Because we assume that observed values are close to the truth value, they should have larger probabilities than unobserved values. 
Thus, the unobserved values have lower impact, and we integrate over the observed value space instead of the truth value space:
\begin{align}\label{eq:observe_p}
    \Pr[v\,|\,e,a,s] & \approx \frac{\sum_{v^*\in V^*_{e,a}} \Pr[v\,|\,v^*, s] \Pr[v^*\,|\,e,a]}{\sum_{v^*\in V^*_{e,a}}\Pr[v^*\,|\,e,a]},
\end{align}
where $V^*_{e,a}$ is the possible truth value set for entity $e$ and attribute $a$. 
Note that $\sum_{v^*\in V^*_{e,a}}\Pr[v^*\,|\,e,a]$ is used to normalize the conditional probability such that it falls into $[0, 1]$. 
To evaluate the above observed value probability, we use the confusion probability to obtain $\Pr[v\,|\,v^*, s]$ and the fact scoring function $\mathcal{F}$ to obtain $\Pr[v^*\,|\,e,a]=\exp(-\mathcal{F}(e,a,v))$. 
For literal values, $\Pr[v^*\,|\,e,a]=\mathcal{F}(e, a, v^*)$, and for entities, $\Pr[v^*\,|\,e,a]=\mathcal{F}(e, a, \operatorname{LE-ANN}(v^*))$.

Based on the observed value probability, we define the loss function for claims as the negative logarithm likelihood function of all claims:
\begin{align}\label{eq:claim_loss}
  \mathcal{L}_{claim} = -\sum_{(e,a,v,s)\in C}\log \Pr[v\,|\,e,a,s],
\end{align}
where $C$ is the claim set.

\subsection{Semi-supervised Learning}

We train our model and discover the truths in a semi-supervised manner. 
For truth inference, each fact is the label of corresponding claims. 
As consistent claims help determine the truths, we combine labeled and unlabeled claims together to train our model. 
Moreover, as claims may contain noises but facts may not, we first train the fact scoring model with facts, and then jointly train the fact scoring model and the truth inference model with facts and claims.

As shown in Algorithm~\ref{algo:ti}, we design a semi-supervised learning algorithm to discover a new fact set $T_{new}$ from a claim set $C$ with the help of existing facts $T$ in the KG. 
In Lines 2--4, we learn the parameters of fact scoring model up to $FACT\_EPOCHS$ epochs. 
In Lines 5--12, we learn the parameters of both fact scoring model and truth inference model up to $INFERENCE\_EPOCHS$ epochs. 
In Lines 7--11, we traverse each claim $(e, a, v, s)\in C$ and compute the observed value probability. 
For each claim $(e, a, v, s)$, we first find all possible truth value set $V^*_{e,a}$ (Line 8), then compute the confusion probability $\Pr[v\,|\,v^*,s]$ with the value alignment model and fact plausibility $\Pr[v^*\,|\,e,a]$ with the fact scoring model (Lines 9--10), and finally update the claim loss with the observed value probability $\Pr[v\,|\,e,a,s]$ using Eq.~(\ref{eq:observe_p}) (Line 11).
In Lines 13--14, for each claim $(e, a, v, s)$, we add $(e, a, v)$ in the new fact set $T_{new}$ if the fact plausibility of $(e,a,v)$ exceeds the threshold $0.5$.

The time complexity of Algorithm~\ref{algo:ti} is $O\big(FACT\_EPOCHS\,|T| + INFERENCE\_EPOCHS\,|C|\big)$.
The total parameter number of our method is  $(K-1) d_h^2 + d_h d_e + d_e(\frac{1}{4}d_e+\frac{1}{2}|E|+d_{BERT}+N_{rel}+N_{lit}) + N_{lit} + d_{ll} (d_{BERT}+ 1) + 8 d_{BERT}^2 + d_{le}(d_{BERT}+d_e) + N_{lit} + N_{rel} + |S|+ d_e d_{de}$.
See Appendix~\ref{app:comp} for more details.


\begin{algorithm}[t]
\KwIn{Fact set $T=T_{lit}\cup T_{rel}$ and claim set $C$}
\KwOut{New fact set $T_{new}$}
Initialize model parameters, and let $T_{new}\leftarrow\emptyset$\;
\For{$epoch=1,2,\dots,FACT\_EPOCHS$}{
	Compute $\mathcal{L}_{fact}$ based on Eq.~(\ref{eq:fact_loss})\;
	Update model parameters based on $\mathcal{L}_{fact}$\;
}
\For{$epoch=1,2,\dots,INFERENCE\_EPOCHS$}{
	Compute $\mathcal{L}_{fact}$ based on Eq.~(\ref{eq:fact_loss}), and let $\mathcal{L}_{claim}\leftarrow 0$\;
	\ForEach(\tcp*[h]{Compute $\mathcal{L}_{claim}$}){$(e, a, v, s)\in C$}{
		$V^*_{e,a}\leftarrow\{v^*\,|\,(e,a, v^*,s')\in C\}$\;
		\ForEach{$v^*\in V^*_{e,a}$}{Compute $\Pr[v\,|\,v^*, s]$ and $\Pr[v^*\,|\,e, a]$\;}
		$\mathcal{L}_{claim}\leftarrow \mathcal{L}_{claim} - \log \Pr[v\,|\, e, a, s]$\;
	}
	Update model parameters based on $\mathcal{L}_{fact} + \mathcal{L}_{claim}$\;
}
\ForEach{$(e, a, v, s)\in C$}{
	\lIf{$\Pr[v\,|\,e,a] > 0.5$}{$T_{new}\leftarrow T_{new}\cup \{(e, a ,v)\}$}
}
\Return $T_{new}$\;
\caption{Semi-supervised truth inference algorithm}
\label{algo:ti}
\end{algorithm}

\section{Evaluation}
\label{sec:exp}

In this section, we assess the proposed TKGC and report our experimental results.
Dataset and source code are accessible online (\url{https://github.com/nju-websoft/TKGC}).

\smallskip 
\noindent\textbf{Hyperparameters.} We implement TKGC on a server with 4 CPUs, 32GB memory and a NVIDIA Tesla V100 graphics card. 
Due to the space limitation, we report the hyperparameters in Appendix~\ref{app:hp}.

\smallskip
\noindent\textbf{Datasets.} We choose a recent open KG completion dataset created by OKELE~\cite{OKELE} to evaluate the performance of TKGC, because this dataset contains noisy facts crawled from web pages and manually-labeled true facts.
Table~\ref{tab:exp_dataset} depicts its statistical data.
This dataset contains 10 popular classes of entities in Freebase~\cite{Freebase}, and each class contains 1,000 entities for training, 100 entities for validation and 100 entities for testing. 
In total, it contains 191,759 facts. 
We reuse these 12,000 entities as seeds, and collect their one-hop and two-hop facts that do not appear in the test set to construct a subgraph of Freebase, which serves as the prior knowledge for TKGC.

\begin{table}
\centering
\caption{Dataset statistics}
\label{tab:exp_dataset}
{\small
\begin{tabular}{l|cc|l|cc}
  \toprule
    \multirow{2}{*}{Classes} & \multicolumn{2}{c|}{Facts} & \multirow{2}{*}{Classes} & \multicolumn{2}{c}{Facts} \\
    \cmidrule{2-3}\cmidrule{5-6}
    & Literal & Relational & & Literal & Relational \\
  \midrule
    actor   & \ \, 330 & 64,983 & album & 155 & \ \ 5,897\\
    book    & \ \, 499 & 10,776 &  building & 361 & \ \ 2,823 \\
    drug    & 1,002 & 26,432 & film & 576 & 45,233 \\
    food    & \ \, 842 & 23,041 & mountain & 623 & \ \ 2,720 \\
    ship    & \ \, 852 & \ \ 1,805 & software & 487 & \ \ 2,322 \\
  \bottomrule
\end{tabular}}
\end{table}

\smallskip
\noindent\textbf{Evaluation metrics.} 
Following \cite{MTKGNN,KBLRN}, we use precision (P), recall (R) and F1-score as the metrics for relational facts; and mean absolute error (MAE) and root of mean square error (RMSE) as the metrics for literal facts. 

\subsection{Overall Performance}

\noindent\textbf{Competitors.} For comparison, we choose 10 competing methods from four categories: 
(i) Four conventional KG completion methods only for relational facts, namely TransE~\cite{TransE}, BoxE~\cite{BoxE}, DualE~\cite{DualE}, ConEx~\cite{ConvEX} and M$^2$GNN~\cite{M2KGNN}, in which BoxE, DualE, ConEx and M$^2$GNN are the state-of-the-arts achieving leading performance on benchmark datasets. 
In this experiment, we force them to treat literals as entities. 
(ii) Two KG completion methods for both relational and literal facts, MT-KGNN~\cite{MTKGNN} and KBLRN~\cite{KBLRN}. 
(iii) Two open KG completion methods supporting external texts, OWE~\cite{OWE} and ConMask~\cite{ConMask}. 
We adopt the TransE version of OWE in this experiment. 
(iv) Two open KG completion methods addressing noisy claims, Knowledge Vault~\cite{KnowledgeVault} and OKELE~\cite{OKELE}.
Note that, for all of them, we carefully read their papers and tune the (hyper)parameters.

\begin{table}
\centering
\caption{Overall performance}
\label{tab:exp_overall}
{\small
\begin{tabular}{l|cc|ccc}
  \toprule
    Methods & MAE$\,\downarrow$ & RMSE$\,\downarrow$ & P$\,\uparrow$ & R$\,\uparrow$ & F1$\,\uparrow$ \\
  \midrule
    TransE          & 0.270 & 0.289 & 0.259 & 0.327 & 0.289 \\
    BoxE            & 0.183 & 0.192 & 0.343 & 0.361 & 0.351 \\
    DualE           & 0.162 & 0.173 & 0.379 & 0.354 & 0.366 \\
    ConEX           & 0.154 & 0.164 & 0.399 & 0.346 & 0.371 \\
    M$^2$GNN        & 0.167 & 0.175 & 0.393 & 0.372 & 0.382 \\
  \midrule
    MT-KGNN         & 0.096 & 0.105 & 0.377 & 0.354 & 0.365 \\
    KBLRN           & 0.109 & 0.116 & 0.323 & 0.322 & 0.322 \\
  \midrule
    OWE (TransE)    & 0.113 & 0.118 & 0.351 & 0.421 & 0.383 \\
    ConMask         & 0.094 & 0.102 & 0.376 & 0.443 & 0.407 \\
  \midrule
    Knowledge Vault & 0.115 & 0.123 & 0.385 & 0.455 & 0.417 \\
    OKELE           & 0.078 & 0.082 & 0.436 & 0.485 & 0.459 \\
  \midrule
    TKGC (ours)     & \textbf{0.054} & \textbf{0.062} & \textbf{0.524} & \textbf{0.491} & \textbf{0.507} \\
  \bottomrule
\end{tabular}}
\end{table}

\smallskip
\noindent\textbf{Results.} Table~\ref{tab:exp_overall} shows the comparison results, and we obtain the following findings: 
For literal fact completion, (i) TransE, BoxE, DualE, ConvEX and M$^2$KGNN perform worst, due to that these four methods only focus on modeling relational facts.
The other methods with specific design of literal fact scoring functions can deal with the literal fact completion better. 
(ii) TKGC achieves the best performance.
Compared with the second-best method OKELE, TKGC improves 0.024 of MAE and 0.020 of RMSE. 
The key reason is that TKGC additionally considers the correlation with existing facts in the KG to enhance truth inference.
(iii) The remaining five methods, namely MT-KGNN, KBLRN, OWE, ConMask and Knowledge Vault, obtain comparable results.
Each method has its own pros and cons. 
MT-KGNN and KBLRN leverage dedicated fact scoring functions for numbers, but they cannot handle entities outside the KG. 
OWE, ConMask and Knowledge Vault do not process numbers separately, but they can leverage external knowledge.

For relational fact completion, (i) compared with TransE, BoxE, DualE, ConEX, M$^2$GNN, MT-KGNN and KBLRN, the open KG completion methods, including OWE, ConMask, Knowledge Vault, OKELE and TKGC, increase at least 0.049 of recall. 
Those methods without external knowledge just choose entities inside the KG for completion, which narrows the range of candidate entities. 
In contrast, the open KG completion methods can leverage the external knowledge from web pages to discover new facts with entities outside the KG. 
(ii) Compared with OWE and ConMask, Knowledge Vault, OKELE and TKGC achieve better precision. 
The reason is that OWE and ConMask neglect the noises in claims, while Knowledge Vault, OKELE and TKGC cope with the noises and eliminate incorrect ones. 
(iii) Compared with the best competitor OKELE, TKGC improves 0.088 of precision, 0.006 of recall and 0.048 of F1-score. 
The root cause is that TKGC not only leverages facts in the KG to improve truth inference, but also makes subtle comparisons with entities based on value alignment. 
Differently, OKELE ignores the entity and attribute information when validating the values in claims, and its pipeline workflow causes more errors.

See Appendix~\ref{app:rt} for the runtime comparison.

\subsection{Evaluation of Fact Scoring}

To evaluate the effectiveness of fact scoring functions for different types of values, we modify two variants of TKGC, namely TKGC without literal fact scoring function (w/o $\mathcal{F}_{lit}$) and TKGC without relational fact scoring function (w/o $\mathcal{F}_{rel}$). 
We compare their performance on KG completion to the full TKGC. 
Notice that TKGC w/o $\mathcal{F}_{lit}$ cannot complete missing literal values (marked as ``N/A''), thus we only test its performance on relational facts. 
Similarly, we only test the performance of TKGC w/o $\mathcal{F}_{rel}$ on completing missing literal facts.
This experiment can be regarded as an ablation study.

As shown in Table~\ref{tab:exp_score}, we have the following finding: 
TKGC w/o $\mathcal{F}_{lit}$ loses 0.099 of precision, 0.010 of recall and 0.056 of F1-score when completing relational facts, and TKGC w/o $\mathcal{F}_{rel}$ increases 0.017 of MAE and 0.021 of RMSE. 
These significant performance drops show that literal fact scoring and relational fact scoring both contribute to the overall performance and can benefit each other. 
For example, \textit{Titanic} and \textit{Britannic} have the same \textit{overall length} ``269m'', as both of their \textit{ship types} are \textit{Olympic Class Ocean Liner}.

\begin{table}
\centering
\caption{Ablation study of fact scoring functions}
\label{tab:exp_score}
{\small
\begin{tabular}{l|cc|ccc}
  \toprule & MAE$\,\downarrow$ & RMSE$\,\downarrow$ & P$\,\uparrow$ & R$\,\uparrow$ & F1$\,\uparrow$ \\
  \midrule TKGC w/o $\mathcal{F}_{lit}$ & N/A & N/A & 0.425 & 0.481 & 0.451 \\
           TKGC w/o $\mathcal{F}_{rel}$ & 0.071 & 0.083 & N/A & N/A & N/A \\
           TKGC (full) & 0.054 & 0.062 & 0.524 & 0.491 & 0.507 \\
    \bottomrule
\end{tabular}}
\end{table}

\begin{table}
\centering
\caption{Comparison of literal-literal alignment methods}
\label{tab:exp_lla}
{\small
\begin{tabular}{l|ccc}
  \toprule
    Methods & P$\,\uparrow$ & R$\,\uparrow$ & F1$\,\uparrow$ \\
  \midrule
    Exact matching  & \textbf{0.949} & 0.293 & 0.448 \\
    Edit distance   & 0.691 & 0.721 & 0.706 \\
  \midrule
    ESIM            & 0.718 & 0.728 & 0.723 \\
    RE2             & 0.792 & 0.755 & 0.773 \\
  \midrule
    DeepMatcher     & 0.814 & 0.782 & 0.798 \\
    EMTransformer   & 0.890 & 0.783 & 0.833 \\
  \midrule
    TKGC            & 0.929 & \textbf{0.813} & \textbf{0.867} \\
  \bottomrule
\end{tabular}}
\end{table}

\subsection{Evaluation of Value Alignment}

\noindent\textbf{Effectiveness of literal-literal alignment.} We test the literal-literal alignment network with several string matching methods. 
We manually label each pair of values in the dataset corresponding to the same attribute of the same entity, and create a literal-literal alignment dataset containing 1,126 literal pairs and 298 matches. 
We split the dataset into three parts with the ratio of 3:1:1 for training, validation and evaluation, respectively.
In this experiment, we choose six competitors from three categories: 
(i) Two conventional string similarity measures~\cite{DIBook}, exact matching and Levenshtein edit distance. 
(ii) Two deep text matching models, ESIM~\cite{ESIM} and RE2 \cite{RE2}. 
(iii) Two deep entity matching models based on literals, DeepMatcher~\cite{DeepMatcher} and EMTransformer~\cite{EMTransformer}. 
As EMTransformer directly employs BERT, we do not include BERT~\cite{BERT} as a baseline.

Table~\ref{tab:exp_lla} lists the comparison results, and we gain a few observations: 
(i) Exact matching obtains the best precision while the worst recall, simply because it cannot handle different literals expressing the same meaning. 
(ii) The five deep models perform better than the two similarity measures, as they can tolerate heterogeneity with token embeddings. 
(iii) Among these deep models, TKGC achieves the best precision, recall and F1-score.

\begin{table}
\centering
\caption{Comparison of literal-entity alignment methods}
\label{tab:exp_lea}
{\small
\begin{tabular}{l|ccc|ccc}
  \toprule
    \multirow{2}{*}{Methods} & \multicolumn{3}{c|}{Inside KG} & \multicolumn{3}{c}{Outside KG} \\
    \cmidrule{2-4} \cmidrule{5-7}
     & P$\,\uparrow$ & R$\,\uparrow$ & F1$\,\uparrow$ & P$\,\uparrow$ & R$\,\uparrow$ & F1$\,\uparrow$ \\
  \midrule
    OWE             & 0.358 & 0.428 & 0.390 & 0.314 & 0.384 & 0.346 \\
    ConMask         & 0.376 & 0.457 & 0.413 & 0.373 & 0.367 & 0.370 \\
  \midrule
    Knowledge Vault & 0.386 & 0.471 & 0.424 & 0.381 & 0.369 & 0.375 \\
    OKELE           & 0.436 & 0.491 & 0.462 & 0.434 & 0.452 & 0.443 \\
  \midrule
    TKGC            & \textbf{0.537} & \textbf{0.492} & \textbf{0.514} & \textbf{0.461} & \textbf{0.483} & \textbf{0.471} \\
  \bottomrule
\end{tabular}}
\end{table}

\smallskip
\noindent\textbf{Effectiveness of literal-entity alignment.} The target of literal-entity alignment is to validate the correctness of facts with corresponding entities outside the KG. 
By feeding the entity embeddings produced by literal-entity alignment into the fact scoring function, we evaluate the performance of TKGC on completing relational facts with entities inside and outside the KG.

Table~\ref{tab:exp_lea} presents the results, and we have some findings: 
(i) All the methods achieve higher precision, recall and F1-score on completing the facts with entities inside the KG than those outside the KG. 
This shows the difficulty of KG completion outside the KG.
(ii) For finding the facts with entities outside the KG, TKGC gains the best performance. 
However, this is still a challenging problem and worth further research.

\begin{figure*}
  \centering
  \subfigure[Noises]{
    \label{fig:noise}
    \includegraphics[width=.45\columnwidth]{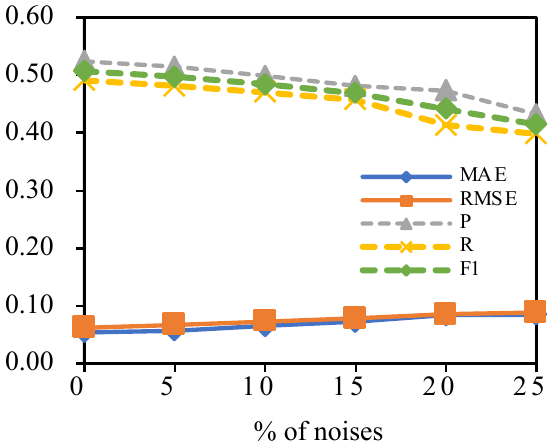}
  }
  \subfigure[Claims]{
    \label{fig:claim}
    \includegraphics[width=.45\columnwidth]{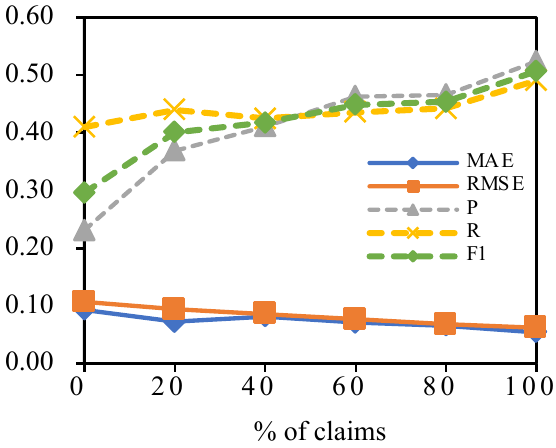}
  }
  \subfigure[Prior knowledge]{
    \label{fig:prior}
    \includegraphics[width=.45\columnwidth]{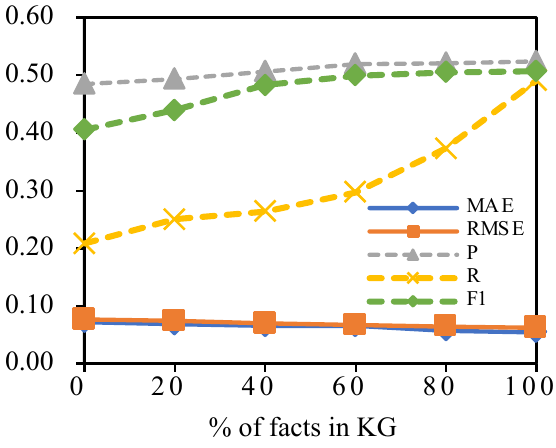}
  }
  \subfigure[Domains]{
    \label{fig:domain}
    \includegraphics[width=.54\columnwidth]{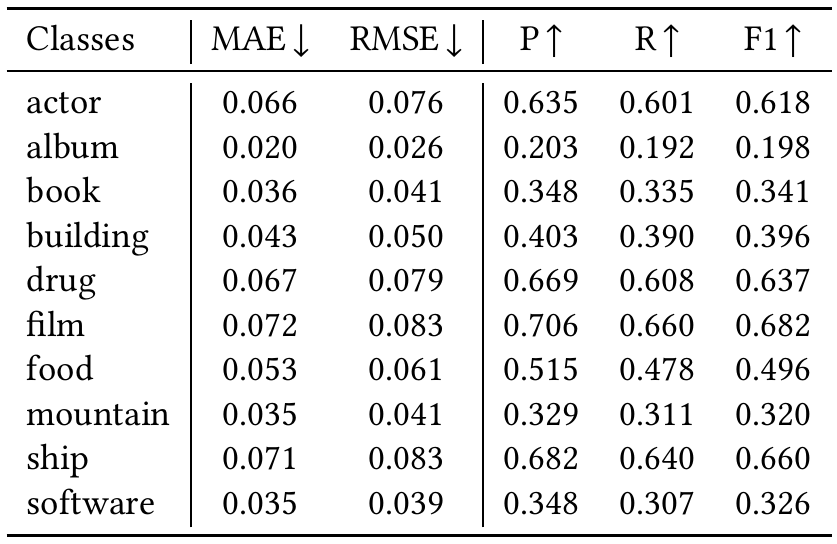}
  }
  \caption{Sensitivity analysis of noises, claims, prior knowledge and domains}
  \label{fig:sensitivity}
\end{figure*}

\subsection{Evaluation of Truth Inference}

We choose eight popular truth inference methods from four categories \cite{TDSurvey,TISurvery}: 
(i) a direct computation method, majority voting~\cite{TISurvery}, 
(ii) two iterative methods, TruthFinder \cite{TruthFinder} and PooledInvestment \cite{PooledInvestment}, 
(iii) an optimization-based method, CATD \cite{CATD}, and 
(iv) four probabilistic graphical models, LTM \cite{LTM}, LCA \cite{LCA}, MBM \cite{MBM} and BWA \cite{BWA}.
It is worth mentioning that majority voting, TruthFinder, PooledInvestment, LTM, LCA and MBM only address single-choice truth inference. 
Following \cite{CATD,TISurvery}, we adapt them to conduct truth inference on multi-valued attributes for a fair comparison.

As shown in Table~\ref{tab:exp_ti}, we get the following findings: 
(i) For literal facts, compared to the best competitor LTM, TKGC reduces 0.017 of MAE and 0.031 of RMSE. 
The main reason is that TKGC leverages correlation with existing facts to determine literal values. 
For example, \textit{Mount Everest} and \textit{Nanga Parbat} have similar \textit{elevations}, as they are both located in the \textit{Himalayas} mountain range.
(ii) For relational facts, compared with the best competitor CATD, TKGC improves 0.092 of precision, 0.068 of recall and 0.080 of F1-score. 
One reason is that TKGC can leverage facts in the KG to judge the correctness of claims. 
Another reason is that all competitors process entities independently, but TKGC makes subtle comparisons between entities, e.g., $(\textit{Born This Way}, \textit{genre}, \textit{Pop Rock})$ can infer $(\textit{Born This Way}, \textit{genre},$ $\textit{Pop})$.

\begin{table}
\centering
\caption{Comparison of truth inference methods}
\label{tab:exp_ti}
{\small
\begin{tabular}{l|cc|ccc}
  \toprule
    Methods & MAE$\,\downarrow$ & RMSE$\,\downarrow$ & P$\,\uparrow$ & R$\,\uparrow$ & F1$\,\uparrow$ \\
  \midrule
    Majority voting     & 0.134 & 0.173 & 0.321 & 0.419 & 0.364 \\
  \midrule
    TruthFinder         & 0.129 & 0.153 & 0.279 & 0.374 & 0.320 \\
    PooledInvestment    & 0.091 & 0.108 & 0.397 & 0.380 & 0.388 \\
  \midrule
    CATD                & 0.127 & 0.145 & 0.432 & 0.423 & 0.427 \\
  \midrule
    LTM                 & 0.071 & 0.093 & 0.262 & 0.394 & 0.315 \\
    LCA                 & 0.106 & 0.130 & 0.364 & 0.404 & 0.383 \\
    MBM                 & 0.104 & 0.125 & 0.340 & \textbf{0.539} & 0.417 \\
    BWA                 & 0.088 & 0.102 & 0.414 & 0.408 & 0.411 \\
  \midrule
    TKGC                & \textbf{0.054} & \textbf{0.062} & \textbf{0.524} & 0.491 & \textbf{0.507} \\ 
  \bottomrule
\end{tabular}}
\end{table}

\subsection{Sensitivity Analysis}


\noindent\textbf{Proportion of noises.} We add random noises into the claim set to investigate how TKGC can tolerate noisy data. 
For each value in the claim set, we randomly substitute it by another value with the probability of 0\%, 5\%, 10\%, 15\%, 20\% and 25\%. 
We repeat this experiment five times, and report the results on average. 
We see in Figure~\ref{fig:noise} that, while the noises increase up to 25\%, the RMSE only increases 0.027 and the F1-score drops 0.092, indicating that TKGC can infer correct values from noisy claims to varying degrees.

\smallskip
\noindent\textbf{Proportion of claims.} We analyze how the proportion of claims affects TKGC. 
We randomly preserve 0\%, 20\%, 40\%, 60\%, 80\% and 100\% of claims, and evaluate the performance. 
We repeat this experiment five times, and present the results on average. 
Based on Figure~\ref{fig:claim}, we obtain two findings: 
(i) Compared TKGC with all claims, TKGC without claims increases 0.031 of RMSE, but significantly drops 0.293 of precision and 0.211 of F1-score. 
The main reason is that claims can provide facts with unseen entities.
(ii) As the amount of claims increases, the performance of TKGC consistently improves. 
One important reason is that the truth inference in TKGC improves its accuracy based on redundancy. 
When there are not enough claims, TKGC cannot rule out wrong facts. 

\smallskip
\noindent\textbf{Proportion of prior knowledge.} We also analyze how the proportion of facts in the KG affects TKGC. 
We randomly preserve 0\%, 20\%, 40\%, 60\%, 80\% and 100\% of facts in the KG, and evaluate the performance. 
Again, we repeat this experiment five times, and report the results on average. 
As shown in Figure~\ref{fig:prior}, we receive two observations: 
(i) When there is no prior knowledge, TKGC only increases 0.015 of RMSE, but significantly loses 0.283 of recall and 0.102 of F1-score. 
The main reason is that TKGC cannot learn entity embeddings to capture related entities without existing facts in the KG, thus it fails to find these entities. 
(ii) As the proportion of prior knowledge increases, the performance of TKGC largely improves. 
This shows that the embeddings learned by prior knowledge can improve the quality of truth inference.

\smallskip
\noindent\textbf{Domain variance.} We present the performance variance of TKGC on different domains in Figure~\ref{fig:domain}. 
We observe that the F1-score of TKGC varies from 0.198 to 0.682, and the standard deviation of F1-score is 0.173.
The main factor is the quality of data sources. 
For example, TKGC obtains the highest F1-score on the ``film'' class, which includes some high-quality  sources like IMDB and Metacritic.

\section{Conclusion}
\label{sec:concl}
In this paper, we propose a trustworthy KG completion method based on multi-sourced noisy data. 
We introduce a holistic scoring function that measures the plausibility of both relational facts and literal facts with various value types.
We design value alignment networks to resolve the heterogeneous values in claims and predict some of them as entities.
We propose a truth inference model to incorporate data source qualities into the fact scoring model and infer the truths from inconsistent values by semi-supervised learning.
The experimental results show that our method consistently achieves the best performance.
Compared with the best competitors for KG completion, our method reduces 0.020 of RMSE for completing literal facts, and improves 0.048 of F1-score for completing relational facts. 
Our models for value alignment, fact scoring with unseen entities and truth inference also gain superior results.
In future work, we plan to extend our method to choose data sources for claim retrieval.
We also want to apply our method to validate the facts from relation extraction.

\begin{acks}
This work is supported by National Natural Science Foundation of China (No. 61872172), and Alibaba Group through Alibaba Research Fellowship Program.
\end{acks}

\bibliographystyle{ACM-Reference-Format}
\bibliography{main}


\begin{thebibliography}{51}


\ifx \showCODEN    \undefined \def \showCODEN     #1{\unskip}     \fi
\ifx \showDOI      \undefined \def \showDOI       #1{#1}\fi
\ifx \showISBNx    \undefined \def \showISBNx     #1{\unskip}     \fi
\ifx \showISBNxiii \undefined \def \showISBNxiii  #1{\unskip}     \fi
\ifx \showISSN     \undefined \def \showISSN      #1{\unskip}     \fi
\ifx \showLCCN     \undefined \def \showLCCN      #1{\unskip}     \fi
\ifx \shownote     \undefined \def \shownote      #1{#1}          \fi
\ifx \showarticletitle \undefined \def \showarticletitle #1{#1}   \fi
\ifx \showURL      \undefined \def \showURL       {\relax}        \fi
\providecommand\bibfield[2]{#2}
\providecommand\bibinfo[2]{#2}
\providecommand\natexlab[1]{#1}
\providecommand\showeprint[2][]{arXiv:#2}

\bibitem[\protect\citeauthoryear{Abboud, Ceylan, Lukasiewicz, and
  Salvatori}{Abboud et~al\mbox{.}}{2020}]%
        {BoxE}
\bibfield{author}{\bibinfo{person}{Ralph Abboud}, \bibinfo{person}{Ismail
  Ceylan}, \bibinfo{person}{Thomas Lukasiewicz}, {and} \bibinfo{person}{Tommaso
  Salvatori}.} \bibinfo{year}{2020}\natexlab{}.
\newblock \showarticletitle{{BoxE}: {A} Box Embedding Model for Knowledge Base
  Completion}. In \bibinfo{booktitle}{\emph{NeurIPS}}.
  \bibinfo{publisher}{Curran Associates, Inc.}, \bibinfo{address}{online},
  \bibinfo{pages}{9649--9661}.
\newblock


\bibitem[\protect\citeauthoryear{Auer, Bizer, Kobilarov, Lehmann, Cyganiak, and
  Ives}{Auer et~al\mbox{.}}{2007}]%
        {DBpedia}
\bibfield{author}{\bibinfo{person}{S{\"{o}}ren Auer},
  \bibinfo{person}{Christian Bizer}, \bibinfo{person}{Georgi Kobilarov},
  \bibinfo{person}{Jens Lehmann}, \bibinfo{person}{Richard Cyganiak}, {and}
  \bibinfo{person}{Zachary~G. Ives}.} \bibinfo{year}{2007}\natexlab{}.
\newblock \showarticletitle{{DBpedia}: {A} Nucleus for a Web of Open Data}. In
  \bibinfo{booktitle}{\emph{ISWC/ASWC}}. \bibinfo{publisher}{Springer},
  \bibinfo{address}{Busan, South Korea}, \bibinfo{pages}{722--735}.
\newblock


\bibitem[\protect\citeauthoryear{Balazevic, Allen, and Hospedales}{Balazevic
  et~al\mbox{.}}{2019}]%
        {TuckER}
\bibfield{author}{\bibinfo{person}{Ivana Balazevic}, \bibinfo{person}{Carl
  Allen}, {and} \bibinfo{person}{Timothy~M. Hospedales}.}
  \bibinfo{year}{2019}\natexlab{}.
\newblock \showarticletitle{{TuckER}: Tensor Factorization for Knowledge Graph
  Completion}. In \bibinfo{booktitle}{\emph{EMNLP-IJCNLP}}.
  \bibinfo{publisher}{ACL}, \bibinfo{address}{Hong Kong, China},
  \bibinfo{pages}{5184--5193}.
\newblock


\bibitem[\protect\citeauthoryear{Bollacker, Evans, Paritosh, Sturge, and
  Taylor}{Bollacker et~al\mbox{.}}{2008}]%
        {Freebase}
\bibfield{author}{\bibinfo{person}{Kurt~D. Bollacker}, \bibinfo{person}{Colin
  Evans}, \bibinfo{person}{Praveen Paritosh}, \bibinfo{person}{Tim Sturge},
  {and} \bibinfo{person}{Jamie Taylor}.} \bibinfo{year}{2008}\natexlab{}.
\newblock \showarticletitle{Freebase: {A} Collaboratively Created Graph
  Database for Structuring Human Knowledge}. In
  \bibinfo{booktitle}{\emph{SIGMOD}}. \bibinfo{publisher}{ACM},
  \bibinfo{address}{Vancouver, BC, Canada}, \bibinfo{pages}{1247--1250}.
\newblock


\bibitem[\protect\citeauthoryear{Bordes, Usunier, Garc{\'{\i}}a{-}Dur{\'{a}}n,
  Weston, and Yakhnenko}{Bordes et~al\mbox{.}}{2013}]%
        {TransE}
\bibfield{author}{\bibinfo{person}{Antoine Bordes}, \bibinfo{person}{Nicolas
  Usunier}, \bibinfo{person}{Alberto Garc{\'{\i}}a{-}Dur{\'{a}}n},
  \bibinfo{person}{Jason Weston}, {and} \bibinfo{person}{Oksana Yakhnenko}.}
  \bibinfo{year}{2013}\natexlab{}.
\newblock \showarticletitle{Translating Embeddings for Modeling
  Multi-relational Data}. In \bibinfo{booktitle}{\emph{NIPS}}.
  \bibinfo{publisher}{Curran Associates, Inc.}, \bibinfo{address}{Lake Tahoe,
  NV, USA}, \bibinfo{pages}{2787--2795}.
\newblock


\bibitem[\protect\citeauthoryear{Brunner and Stockinger}{Brunner and
  Stockinger}{2020}]%
        {EMTransformer}
\bibfield{author}{\bibinfo{person}{Ursin Brunner} {and} \bibinfo{person}{Kurt
  Stockinger}.} \bibinfo{year}{2020}\natexlab{}.
\newblock \showarticletitle{Entity Matching with {Transformer} Architectures -
  {A} Step Forward in Data Integration}. In \bibinfo{booktitle}{\emph{EDBT}}.
  \bibinfo{publisher}{OpenProceedings.org}, \bibinfo{address}{Copenhagen,
  Denmark}, \bibinfo{pages}{463--473}.
\newblock


\bibitem[\protect\citeauthoryear{Cao, Wang, Huang, and Hu}{Cao
  et~al\mbox{.}}{2020}]%
        {OKELE}
\bibfield{author}{\bibinfo{person}{Ermei Cao}, \bibinfo{person}{Difeng Wang},
  \bibinfo{person}{Jiacheng Huang}, {and} \bibinfo{person}{Wei Hu}.}
  \bibinfo{year}{2020}\natexlab{}.
\newblock \showarticletitle{Open Knowledge Enrichment for Long-tail Entities}.
  In \bibinfo{booktitle}{\emph{WWW}}. \bibinfo{publisher}{ACM},
  \bibinfo{address}{Taipei, Taiwan}, \bibinfo{pages}{384--394}.
\newblock


\bibitem[\protect\citeauthoryear{Cao, Xu, Yang, Cao, and Huang}{Cao
  et~al\mbox{.}}{2021}]%
        {DualE}
\bibfield{author}{\bibinfo{person}{Zongsheng Cao}, \bibinfo{person}{Qianqian
  Xu}, \bibinfo{person}{Zhiyong Yang}, \bibinfo{person}{Xiaochun Cao}, {and}
  \bibinfo{person}{Qingming Huang}.} \bibinfo{year}{2021}\natexlab{}.
\newblock \showarticletitle{Dual Quaternion Knowledge Graph Embeddings}. In
  \bibinfo{booktitle}{\emph{AAAI}}. \bibinfo{publisher}{AAAI Press},
  \bibinfo{address}{online}, \bibinfo{pages}{6894--6902}.
\newblock


\bibitem[\protect\citeauthoryear{Chen, Zhu, Ling, Wei, Jiang, and Inkpen}{Chen
  et~al\mbox{.}}{2017}]%
        {ESIM}
\bibfield{author}{\bibinfo{person}{Qian Chen}, \bibinfo{person}{Xiaodan Zhu},
  \bibinfo{person}{Zhen{-}Hua Ling}, \bibinfo{person}{Si Wei},
  \bibinfo{person}{Hui Jiang}, {and} \bibinfo{person}{Diana Inkpen}.}
  \bibinfo{year}{2017}\natexlab{}.
\newblock \showarticletitle{Enhanced {LSTM} for Natural Language Inference}. In
  \bibinfo{booktitle}{\emph{ACL}}. \bibinfo{publisher}{ACL},
  \bibinfo{address}{Vancouver, BC, Canada}, \bibinfo{pages}{1657--1668}.
\newblock


\bibitem[\protect\citeauthoryear{Demir and Ngomo}{Demir and Ngomo}{2021}]%
        {ConvEX}
\bibfield{author}{\bibinfo{person}{Caglar Demir} {and}
  \bibinfo{person}{Axel{-}Cyrille~Ngonga Ngomo}.}
  \bibinfo{year}{2021}\natexlab{}.
\newblock \showarticletitle{Convolutional Complex Knowledge Graph Embeddings}.
  In \bibinfo{booktitle}{\emph{ESWC}}. \bibinfo{publisher}{Springer},
  \bibinfo{address}{Heraklion, Greece}, \bibinfo{pages}{409--424}.
\newblock


\bibitem[\protect\citeauthoryear{Dettmers, Minervini, Stenetorp, and
  Riedel}{Dettmers et~al\mbox{.}}{2018}]%
        {ConvE}
\bibfield{author}{\bibinfo{person}{Tim Dettmers}, \bibinfo{person}{Pasquale
  Minervini}, \bibinfo{person}{Pontus Stenetorp}, {and}
  \bibinfo{person}{Sebastian Riedel}.} \bibinfo{year}{2018}\natexlab{}.
\newblock \showarticletitle{Convolutional {2D} Knowledge Graph Embeddings}. In
  \bibinfo{booktitle}{\emph{AAAI}}. \bibinfo{publisher}{AAAI Press},
  \bibinfo{address}{New Orleans, LA, USA}, \bibinfo{pages}{1811--1818}.
\newblock


\bibitem[\protect\citeauthoryear{Devlin, Chang, Lee, and Toutanova}{Devlin
  et~al\mbox{.}}{2019}]%
        {BERT}
\bibfield{author}{\bibinfo{person}{Jacob Devlin}, \bibinfo{person}{Ming{-}Wei
  Chang}, \bibinfo{person}{Kenton Lee}, {and} \bibinfo{person}{Kristina
  Toutanova}.} \bibinfo{year}{2019}\natexlab{}.
\newblock \showarticletitle{{BERT}: Pre-training of Deep Bidirectional
  Transformers for Language Understanding}. In
  \bibinfo{booktitle}{\emph{NAACL-HLT}}. \bibinfo{publisher}{ACL},
  \bibinfo{address}{Minneapolis, MN, USA}, \bibinfo{pages}{4171--4186}.
\newblock


\bibitem[\protect\citeauthoryear{Doan, Halevy, and Ives}{Doan
  et~al\mbox{.}}{2012}]%
        {DIBook}
\bibfield{author}{\bibinfo{person}{AnHai Doan}, \bibinfo{person}{Alon~Y.
  Halevy}, {and} \bibinfo{person}{Zachary~G. Ives}.}
  \bibinfo{year}{2012}\natexlab{}.
\newblock \bibinfo{booktitle}{\emph{Principles of Data Integration}}.
\newblock \bibinfo{publisher}{Morgan Kaufmann}, \bibinfo{address}{Waltham, MA,
  USA}.
\newblock


\bibitem[\protect\citeauthoryear{Dong, Gabrilovich, Heitz, Horn, Lao, Murphy,
  Strohmann, Sun, and Zhang}{Dong et~al\mbox{.}}{2014}]%
        {KnowledgeVault}
\bibfield{author}{\bibinfo{person}{Xin Dong}, \bibinfo{person}{Evgeniy
  Gabrilovich}, \bibinfo{person}{Geremy Heitz}, \bibinfo{person}{Wilko Horn},
  \bibinfo{person}{Ni Lao}, \bibinfo{person}{Kevin Murphy},
  \bibinfo{person}{Thomas Strohmann}, \bibinfo{person}{Shaohua Sun}, {and}
  \bibinfo{person}{Wei Zhang}.} \bibinfo{year}{2014}\natexlab{}.
\newblock \showarticletitle{{Knowledge Vault}: {A} Web-scale Approach to
  Probabilistic Knowledge Fusion}. In \bibinfo{booktitle}{\emph{KDD}}.
  \bibinfo{publisher}{ACM}, \bibinfo{address}{New York, NY, USA},
  \bibinfo{pages}{601--610}.
\newblock


\bibitem[\protect\citeauthoryear{Garc{\'{\i}}a{-}Dur{\'{a}}n and
  Niepert}{Garc{\'{\i}}a{-}Dur{\'{a}}n and Niepert}{2018}]%
        {KBLRN}
\bibfield{author}{\bibinfo{person}{Alberto Garc{\'{\i}}a{-}Dur{\'{a}}n} {and}
  \bibinfo{person}{Mathias Niepert}.} \bibinfo{year}{2018}\natexlab{}.
\newblock \showarticletitle{{KBLRN}: End-to-end Learning of Knowledge Base
  Representations with Latent, Relational, and Numerical Features}. In
  \bibinfo{booktitle}{\emph{{UAI}}}. \bibinfo{publisher}{AUAI Press},
  \bibinfo{address}{Monterey, CA, USA}, \bibinfo{pages}{372--381}.
\newblock


\bibitem[\protect\citeauthoryear{Guo, Sun, and Hu}{Guo et~al\mbox{.}}{2019}]%
        {RSN}
\bibfield{author}{\bibinfo{person}{Lingbing Guo}, \bibinfo{person}{Zequn Sun},
  {and} \bibinfo{person}{Wei Hu}.} \bibinfo{year}{2019}\natexlab{}.
\newblock \showarticletitle{Learning to Exploit Long-term Relational
  Dependencies in Knowledge Graphs}. In \bibinfo{booktitle}{\emph{ICML}}.
  \bibinfo{publisher}{PMLR}, \bibinfo{address}{Long Beach, CA, USA},
  \bibinfo{pages}{2505--2514}.
\newblock


\bibitem[\protect\citeauthoryear{Hamilton, Ying, and Leskovec}{Hamilton
  et~al\mbox{.}}{2017}]%
        {GraphSage}
\bibfield{author}{\bibinfo{person}{William~L. Hamilton},
  \bibinfo{person}{Zhitao Ying}, {and} \bibinfo{person}{Jure Leskovec}.}
  \bibinfo{year}{2017}\natexlab{}.
\newblock \showarticletitle{Inductive Representation Learning on Large Graphs}.
  In \bibinfo{booktitle}{\emph{NeurIPS}}. \bibinfo{publisher}{Curran
  Associates, Inc.}, \bibinfo{address}{Long Beach, CA, USA},
  \bibinfo{pages}{1024--1034}.
\newblock


\bibitem[\protect\citeauthoryear{Ji, Pan, Cambria, Marttinen, and Yu}{Ji
  et~al\mbox{.}}{2021}]%
        {KGSurvey}
\bibfield{author}{\bibinfo{person}{Shaoxiong Ji}, \bibinfo{person}{Shirui Pan},
  \bibinfo{person}{Erik Cambria}, \bibinfo{person}{Pekka Marttinen}, {and}
  \bibinfo{person}{Philip~S. Yu}.} \bibinfo{year}{2021}\natexlab{}.
\newblock \showarticletitle{A Survey on Knowledge Graphs: Representation,
  Acquisition and Applications}.
\newblock \bibinfo{journal}{\emph{IEEE Transactions on Neural Networks and
  Learning Systems}}  \bibinfo{volume}{early access} (\bibinfo{year}{2021}),
  \bibinfo{pages}{1--21}.
\newblock


\bibitem[\protect\citeauthoryear{Koncel{-}Kedziorski, Bekal, Luan, Lapata, and
  Hajishirzi}{Koncel{-}Kedziorski et~al\mbox{.}}{2019}]%
        {TextGeneration}
\bibfield{author}{\bibinfo{person}{Rik Koncel{-}Kedziorski},
  \bibinfo{person}{Dhanush Bekal}, \bibinfo{person}{Yi Luan},
  \bibinfo{person}{Mirella Lapata}, {and} \bibinfo{person}{Hannaneh
  Hajishirzi}.} \bibinfo{year}{2019}\natexlab{}.
\newblock \showarticletitle{Text Generation from Knowledge Graphs with Graph
  Transformers}. In \bibinfo{booktitle}{\emph{NAACL-HLT}}.
  \bibinfo{publisher}{ACL}, \bibinfo{address}{Minneapolis, MN, USA},
  \bibinfo{pages}{2284--2293}.
\newblock


\bibitem[\protect\citeauthoryear{Kristiadi, Khan, Lukovnikov, Lehmann, and
  Fischer}{Kristiadi et~al\mbox{.}}{2019}]%
        {LiteralE}
\bibfield{author}{\bibinfo{person}{Agustinus Kristiadi},
  \bibinfo{person}{Mohammad~Asif Khan}, \bibinfo{person}{Denis Lukovnikov},
  \bibinfo{person}{Jens Lehmann}, {and} \bibinfo{person}{Asja Fischer}.}
  \bibinfo{year}{2019}\natexlab{}.
\newblock \showarticletitle{Incorporating Literals into Knowledge Graph
  Embeddings}. In \bibinfo{booktitle}{\emph{ISWC}}.
  \bibinfo{publisher}{Springer}, \bibinfo{address}{Auckland, New Zealand},
  \bibinfo{pages}{347--363}.
\newblock


\bibitem[\protect\citeauthoryear{Li, Li, Gao, Su, Zhao, Demirbas, Fan, and
  Han}{Li et~al\mbox{.}}{2014}]%
        {CATD}
\bibfield{author}{\bibinfo{person}{Qi Li}, \bibinfo{person}{Yaliang Li},
  \bibinfo{person}{Jing Gao}, \bibinfo{person}{Lu Su}, \bibinfo{person}{Bo
  Zhao}, \bibinfo{person}{Murat Demirbas}, \bibinfo{person}{Wei Fan}, {and}
  \bibinfo{person}{Jiawei Han}.} \bibinfo{year}{2014}\natexlab{}.
\newblock \showarticletitle{A Confidence-aware Approach for Truth Discovery on
  Long-tail Data}.
\newblock \bibinfo{journal}{\emph{Proceedings of the VLDB Endowment}}
  \bibinfo{volume}{8}, \bibinfo{number}{4} (\bibinfo{year}{2014}),
  \bibinfo{pages}{425--436}.
\newblock


\bibitem[\protect\citeauthoryear{Li, Gao, Meng, Li, Su, Zhao, Fan, and Han}{Li
  et~al\mbox{.}}{2015}]%
        {TDSurvey}
\bibfield{author}{\bibinfo{person}{Yaliang Li}, \bibinfo{person}{Jing Gao},
  \bibinfo{person}{Chuishi Meng}, \bibinfo{person}{Qi Li}, \bibinfo{person}{Lu
  Su}, \bibinfo{person}{Bo Zhao}, \bibinfo{person}{Wei Fan}, {and}
  \bibinfo{person}{Jiawei Han}.} \bibinfo{year}{2015}\natexlab{}.
\newblock \showarticletitle{A Survey on Truth Discovery}.
\newblock \bibinfo{journal}{\emph{ACM SIGKDD Explorations Newsletter}}
  \bibinfo{volume}{17}, \bibinfo{number}{2} (\bibinfo{year}{2015}),
  \bibinfo{pages}{1--16}.
\newblock


\bibitem[\protect\citeauthoryear{Li, Rubinstein, and Cohn}{Li
  et~al\mbox{.}}{2019}]%
        {BWA}
\bibfield{author}{\bibinfo{person}{Yuan Li}, \bibinfo{person}{Benjamin I.~P.
  Rubinstein}, {and} \bibinfo{person}{Trevor Cohn}.}
  \bibinfo{year}{2019}\natexlab{}.
\newblock \showarticletitle{Truth Inference at Scale: {A} Bayesian Model for
  Adjudicating Highly Redundant Crowd Annotations}. In
  \bibinfo{booktitle}{\emph{WWW}}. \bibinfo{publisher}{ACM},
  \bibinfo{address}{San Francisco, CA, USA}, \bibinfo{pages}{1028--1038}.
\newblock


\bibitem[\protect\citeauthoryear{Mudgal, Li, Rekatsinas, Doan, Park, Krishnan,
  Deep, Arcaute, and Raghavendra}{Mudgal et~al\mbox{.}}{2018}]%
        {DeepMatcher}
\bibfield{author}{\bibinfo{person}{Sidharth Mudgal}, \bibinfo{person}{Han Li},
  \bibinfo{person}{Theodoros Rekatsinas}, \bibinfo{person}{AnHai Doan},
  \bibinfo{person}{Youngchoon Park}, \bibinfo{person}{Ganesh Krishnan},
  \bibinfo{person}{Rohit Deep}, \bibinfo{person}{Esteban Arcaute}, {and}
  \bibinfo{person}{Vijay Raghavendra}.} \bibinfo{year}{2018}\natexlab{}.
\newblock \showarticletitle{Deep Learning for Entity Matching: A Design Space
  Exploration}. In \bibinfo{booktitle}{\emph{SIGMOD}}.
  \bibinfo{publisher}{ACM}, \bibinfo{address}{Houston, TX, USA},
  \bibinfo{pages}{19--34}.
\newblock


\bibitem[\protect\citeauthoryear{Nguyen, Vu, Nguyen, Nguyen, and Phung}{Nguyen
  et~al\mbox{.}}{2019}]%
        {CapsE}
\bibfield{author}{\bibinfo{person}{Dai~Quoc Nguyen}, \bibinfo{person}{Thanh
  Vu}, \bibinfo{person}{Tu~Dinh Nguyen}, \bibinfo{person}{Dat~Quoc Nguyen},
  {and} \bibinfo{person}{Dinh~Q. Phung}.} \bibinfo{year}{2019}\natexlab{}.
\newblock \showarticletitle{A Capsule Network-based Embedding Model for
  Knowledge Graph Completion and Search Personalization}. In
  \bibinfo{booktitle}{\emph{NAACL-HLT}}. \bibinfo{publisher}{ACL},
  \bibinfo{address}{Minneapolis, MN, USA}, \bibinfo{pages}{2180--2189}.
\newblock


\bibitem[\protect\citeauthoryear{Niu, Fu, Yang, Li, Chen, Liu, and Zheng}{Niu
  et~al\mbox{.}}{2021}]%
        {MIA}
\bibfield{author}{\bibinfo{person}{Lei Niu}, \bibinfo{person}{Chenpeng Fu},
  \bibinfo{person}{Qiang Yang}, \bibinfo{person}{Zhixu Li},
  \bibinfo{person}{Zhigang Chen}, \bibinfo{person}{Qingsheng Liu}, {and}
  \bibinfo{person}{Kai Zheng}.} \bibinfo{year}{2021}\natexlab{}.
\newblock \showarticletitle{Open-world Knowledge Graph Completion with Multiple
  Interaction Attention}.
\newblock \bibinfo{journal}{\emph{World Wide Web}} \bibinfo{volume}{24},
  \bibinfo{number}{1} (\bibinfo{year}{2021}), \bibinfo{pages}{419--439}.
\newblock


\bibitem[\protect\citeauthoryear{Pasternack and Roth}{Pasternack and
  Roth}{2010}]%
        {PooledInvestment}
\bibfield{author}{\bibinfo{person}{Jeff Pasternack} {and} \bibinfo{person}{Dan
  Roth}.} \bibinfo{year}{2010}\natexlab{}.
\newblock \showarticletitle{Knowing What to Believe (when you already know
  something)}. In \bibinfo{booktitle}{\emph{COLING}}. \bibinfo{publisher}{ACL},
  \bibinfo{address}{Beijing, China}, \bibinfo{pages}{877--885}.
\newblock


\bibitem[\protect\citeauthoryear{Pasternack and Roth}{Pasternack and
  Roth}{2013}]%
        {LCA}
\bibfield{author}{\bibinfo{person}{Jeff Pasternack} {and} \bibinfo{person}{Dan
  Roth}.} \bibinfo{year}{2013}\natexlab{}.
\newblock \showarticletitle{Latent Credibility Analysis}. In
  \bibinfo{booktitle}{\emph{WWW}}. \bibinfo{publisher}{IW3C2},
  \bibinfo{address}{Rio de Janeiro, Brazil}, \bibinfo{pages}{1009--1020}.
\newblock


\bibitem[\protect\citeauthoryear{Rossi, Barbosa, Firmani, Matinata, and
  Merialdo}{Rossi et~al\mbox{.}}{2021}]%
        {LP1}
\bibfield{author}{\bibinfo{person}{Andrea Rossi}, \bibinfo{person}{Denilson
  Barbosa}, \bibinfo{person}{Donatella Firmani}, \bibinfo{person}{Antonio
  Matinata}, {and} \bibinfo{person}{Paolo Merialdo}.}
  \bibinfo{year}{2021}\natexlab{}.
\newblock \showarticletitle{Knowledge Graph Embedding for Link Prediction: {A}
  Comparative Analysis}.
\newblock \bibinfo{journal}{\emph{ACM Transactions on Knowledge Discovery from
  Data}} \bibinfo{volume}{15}, \bibinfo{number}{2} (\bibinfo{year}{2021}),
  \bibinfo{pages}{14:1--14:49}.
\newblock


\bibitem[\protect\citeauthoryear{Russell and Norvig}{Russell and
  Norvig}{2020}]%
        {AI}
\bibfield{author}{\bibinfo{person}{Stuart Russell} {and} \bibinfo{person}{Peter
  Norvig}.} \bibinfo{year}{2020}\natexlab{}.
\newblock \bibinfo{booktitle}{\emph{Artificial Intelligence: A Modern Approach}
  (\bibinfo{edition}{4th} ed.)}.
\newblock \bibinfo{publisher}{Prentice Hall}, \bibinfo{address}{Hoboken, NJ,
  USA}.
\newblock


\bibitem[\protect\citeauthoryear{Sakor, Mulang, Singh, Shekarpour, Vidal,
  Lehmann, and Auer}{Sakor et~al\mbox{.}}{2019}]%
        {Falcon}
\bibfield{author}{\bibinfo{person}{Ahmad Sakor}, \bibinfo{person}{Isaiah~Onando
  Mulang}, \bibinfo{person}{Kuldeep Singh}, \bibinfo{person}{Saeedeh
  Shekarpour}, \bibinfo{person}{Maria{-}Esther Vidal}, \bibinfo{person}{Jens
  Lehmann}, {and} \bibinfo{person}{S{\"{o}}ren Auer}.}
  \bibinfo{year}{2019}\natexlab{}.
\newblock \showarticletitle{Old is Gold: Linguistic Driven Approach for Entity
  and Relation Linking of Short Text}. In
  \bibinfo{booktitle}{\emph{NAACL-HLT}}. \bibinfo{publisher}{ACL},
  \bibinfo{address}{Minneapolis, MN, USA}, \bibinfo{pages}{2336--2346}.
\newblock


\bibitem[\protect\citeauthoryear{Schlichtkrull, Kipf, Bloem, van~den Berg,
  Titov, and Welling}{Schlichtkrull et~al\mbox{.}}{2018}]%
        {RGCN}
\bibfield{author}{\bibinfo{person}{Michael~Sejr Schlichtkrull},
  \bibinfo{person}{Thomas~N. Kipf}, \bibinfo{person}{Peter Bloem},
  \bibinfo{person}{Rianne van~den Berg}, \bibinfo{person}{Ivan Titov}, {and}
  \bibinfo{person}{Max Welling}.} \bibinfo{year}{2018}\natexlab{}.
\newblock \showarticletitle{Modeling Relational Data with Graph Convolutional
  Networks}. In \bibinfo{booktitle}{\emph{ESWC}}.
  \bibinfo{publisher}{Springer}, \bibinfo{address}{Heraklion, Crete, Greece},
  \bibinfo{pages}{593--607}.
\newblock


\bibitem[\protect\citeauthoryear{Shah, Villmow, Ulges, Schwanecke, and
  Shafait}{Shah et~al\mbox{.}}{2019}]%
        {OWE}
\bibfield{author}{\bibinfo{person}{Haseeb Shah}, \bibinfo{person}{Johannes
  Villmow}, \bibinfo{person}{Adrian Ulges}, \bibinfo{person}{Ulrich
  Schwanecke}, {and} \bibinfo{person}{Faisal Shafait}.}
  \bibinfo{year}{2019}\natexlab{}.
\newblock \showarticletitle{An Open-world Extension to Knowledge Graph
  Completion Models}. In \bibinfo{booktitle}{\emph{AAAI}}.
  \bibinfo{publisher}{{AAAI} Press}, \bibinfo{address}{Honolulu, HI, USA},
  \bibinfo{pages}{3044--3051}.
\newblock


\bibitem[\protect\citeauthoryear{Shang, Tang, Huang, Bi, He, and Zhou}{Shang
  et~al\mbox{.}}{2019}]%
        {SACN}
\bibfield{author}{\bibinfo{person}{Chao Shang}, \bibinfo{person}{Yun Tang},
  \bibinfo{person}{Jing Huang}, \bibinfo{person}{Jinbo Bi},
  \bibinfo{person}{Xiaodong He}, {and} \bibinfo{person}{Bowen Zhou}.}
  \bibinfo{year}{2019}\natexlab{}.
\newblock \showarticletitle{End-to-end Structure-aware Convolutional Networks
  for Knowledge Base Completion}. In \bibinfo{booktitle}{\emph{AAAI}}.
  \bibinfo{publisher}{AAAI Press}, \bibinfo{address}{Honolulu, HI, USA},
  \bibinfo{pages}{3060--3067}.
\newblock


\bibitem[\protect\citeauthoryear{Shi and Weninger}{Shi and Weninger}{2018}]%
        {ConMask}
\bibfield{author}{\bibinfo{person}{Baoxu Shi} {and} \bibinfo{person}{Tim
  Weninger}.} \bibinfo{year}{2018}\natexlab{}.
\newblock \showarticletitle{Open-world Knowledge Graph Completion}. In
  \bibinfo{booktitle}{\emph{AAAI}}. \bibinfo{publisher}{{AAAI} Press},
  \bibinfo{address}{New Orleans, LA, USA}, \bibinfo{pages}{1957--1964}.
\newblock


\bibitem[\protect\citeauthoryear{Sun, Wang, Hu, Chen, Dai, Zhang, and Qu}{Sun
  et~al\mbox{.}}{2020}]%
        {AliNet}
\bibfield{author}{\bibinfo{person}{Zequn Sun}, \bibinfo{person}{Chengming
  Wang}, \bibinfo{person}{Wei Hu}, \bibinfo{person}{Muhao Chen},
  \bibinfo{person}{Jian Dai}, \bibinfo{person}{Wei Zhang}, {and}
  \bibinfo{person}{Yuzhong Qu}.} \bibinfo{year}{2020}\natexlab{}.
\newblock \showarticletitle{Knowledge Graph Alignment Network with Gated
  Multi-hop Neighborhood Aggregation}. In \bibinfo{booktitle}{\emph{{AAAI}}}.
  \bibinfo{publisher}{AAAI Press}, \bibinfo{address}{New York, NY, USA},
  \bibinfo{pages}{222--229}.
\newblock


\bibitem[\protect\citeauthoryear{Tay, Tuan, Phan, and Hui}{Tay
  et~al\mbox{.}}{2017}]%
        {MTKGNN}
\bibfield{author}{\bibinfo{person}{Yi Tay}, \bibinfo{person}{Luu~Anh Tuan},
  \bibinfo{person}{Minh~C. Phan}, {and} \bibinfo{person}{Siu~Cheung Hui}.}
  \bibinfo{year}{2017}\natexlab{}.
\newblock \showarticletitle{Multi-task Neural Network for Non-discrete
  Attribute Prediction in Knowledge Graphs}. In
  \bibinfo{booktitle}{\emph{CIKM}}. \bibinfo{publisher}{{ACM}},
  \bibinfo{address}{Singapore}, \bibinfo{pages}{1029--1038}.
\newblock


\bibitem[\protect\citeauthoryear{Tversky}{Tversky}{1977}]%
        {Asymmetric}
\bibfield{author}{\bibinfo{person}{Amos Tversky}.}
  \bibinfo{year}{1977}\natexlab{}.
\newblock \showarticletitle{Features of Similarity}.
\newblock \bibinfo{journal}{\emph{Psychological Review}} \bibinfo{volume}{84},
  \bibinfo{number}{4} (\bibinfo{year}{1977}), \bibinfo{pages}{327--352}.
\newblock


\bibitem[\protect\citeauthoryear{Vashishth, Sanyal, Nitin, and
  Talukdar}{Vashishth et~al\mbox{.}}{2020}]%
        {CompGCN}
\bibfield{author}{\bibinfo{person}{Shikhar Vashishth}, \bibinfo{person}{Soumya
  Sanyal}, \bibinfo{person}{Vikram Nitin}, {and} \bibinfo{person}{Partha~P.
  Talukdar}.} \bibinfo{year}{2020}\natexlab{}.
\newblock \showarticletitle{Composition-based Multi-relational Graph
  Convolutional Networks}. In \bibinfo{booktitle}{\emph{ICLR}}.
  \bibinfo{publisher}{OpenReview.net}, \bibinfo{address}{Addis Ababa,
  Ethiopia}, \bibinfo{pages}{1--16}.
\newblock


\bibitem[\protect\citeauthoryear{Vrandecic and Kr{\"{o}}tzsch}{Vrandecic and
  Kr{\"{o}}tzsch}{2014}]%
        {Wikidata}
\bibfield{author}{\bibinfo{person}{Denny Vrandecic} {and}
  \bibinfo{person}{Markus Kr{\"{o}}tzsch}.} \bibinfo{year}{2014}\natexlab{}.
\newblock \showarticletitle{Wikidata: {A} Free Collaborative Knowledgebase}.
\newblock \bibinfo{journal}{\emph{Commun. ACM}} \bibinfo{volume}{57},
  \bibinfo{number}{10} (\bibinfo{year}{2014}), \bibinfo{pages}{78--85}.
\newblock


\bibitem[\protect\citeauthoryear{Wang, Mao, Wang, and Guo}{Wang
  et~al\mbox{.}}{2017}]%
        {KGESurvey}
\bibfield{author}{\bibinfo{person}{Quan Wang}, \bibinfo{person}{Zhendong Mao},
  \bibinfo{person}{Bin Wang}, {and} \bibinfo{person}{Li Guo}.}
  \bibinfo{year}{2017}\natexlab{}.
\newblock \showarticletitle{Knowledge Graph Embedding: {A} Survey of Approaches
  and Applications}.
\newblock \bibinfo{journal}{\emph{IEEE Transactions on Knowledge and Data
  Engineering}} \bibinfo{volume}{29}, \bibinfo{number}{12}
  (\bibinfo{year}{2017}), \bibinfo{pages}{2724--2743}.
\newblock


\bibitem[\protect\citeauthoryear{Wang, Wei, dos Santos, Wang, Nallapati,
  Arnold, Xiang, Yu, and Cruz}{Wang et~al\mbox{.}}{2021}]%
        {M2KGNN}
\bibfield{author}{\bibinfo{person}{Shen Wang}, \bibinfo{person}{Xiaokai Wei},
  \bibinfo{person}{C{\'{\i}}cero~Nogueira dos Santos}, \bibinfo{person}{Zhiguo
  Wang}, \bibinfo{person}{Ramesh Nallapati}, \bibinfo{person}{Andrew~O.
  Arnold}, \bibinfo{person}{Bing Xiang}, \bibinfo{person}{Philip~S. Yu}, {and}
  \bibinfo{person}{Isabel~F. Cruz}.} \bibinfo{year}{2021}\natexlab{}.
\newblock \showarticletitle{Mixed-curvature Multi-relational Graph Neural
  Network for Knowledge Graph Completion}. In \bibinfo{booktitle}{\emph{WWW}}.
  \bibinfo{publisher}{ACM}, \bibinfo{address}{Ljubljana, Slovenia},
  \bibinfo{pages}{1761--1771}.
\newblock


\bibitem[\protect\citeauthoryear{Wang, Sheng, Fang, Yao, Xu, and Li}{Wang
  et~al\mbox{.}}{2015}]%
        {MBM}
\bibfield{author}{\bibinfo{person}{Xianzhi Wang}, \bibinfo{person}{Quan~Z.
  Sheng}, \bibinfo{person}{Xiu~Susie Fang}, \bibinfo{person}{Lina Yao},
  \bibinfo{person}{Xiaofei Xu}, {and} \bibinfo{person}{Xue Li}.}
  \bibinfo{year}{2015}\natexlab{}.
\newblock \showarticletitle{An Integrated Bayesian Approach for Effective
  Multi-truth Discovery}. In \bibinfo{booktitle}{\emph{CIKM}}.
  \bibinfo{publisher}{ACM}, \bibinfo{address}{Melbourne, Australia},
  \bibinfo{pages}{493--502}.
\newblock


\bibitem[\protect\citeauthoryear{Wu, Petroni, Josifoski, Riedel, and
  Zettlemoyer}{Wu et~al\mbox{.}}{2020}]%
        {BLINK}
\bibfield{author}{\bibinfo{person}{Ledell Wu}, \bibinfo{person}{Fabio Petroni},
  \bibinfo{person}{Martin Josifoski}, \bibinfo{person}{Sebastian Riedel}, {and}
  \bibinfo{person}{Luke Zettlemoyer}.} \bibinfo{year}{2020}\natexlab{}.
\newblock \showarticletitle{Scalable Zero-shot Entity Linking with Dense Entity
  Retrieval}. In \bibinfo{booktitle}{\emph{EMNLP}}. \bibinfo{publisher}{ACL},
  \bibinfo{address}{online}, \bibinfo{pages}{6397--6407}.
\newblock


\bibitem[\protect\citeauthoryear{Wu, Li, Wang, and Zhu}{Wu
  et~al\mbox{.}}{2012}]%
        {Probase}
\bibfield{author}{\bibinfo{person}{Wentao Wu}, \bibinfo{person}{Hongsong Li},
  \bibinfo{person}{Haixun Wang}, {and} \bibinfo{person}{Kenny~Qili Zhu}.}
  \bibinfo{year}{2012}\natexlab{}.
\newblock \showarticletitle{Probase: {A} Probabilistic Taxonomy for Text
  Understanding}. In \bibinfo{booktitle}{\emph{SIGMOD}}.
  \bibinfo{publisher}{ACM}, \bibinfo{address}{Scottsdale, AZ, USA},
  \bibinfo{pages}{481--492}.
\newblock


\bibitem[\protect\citeauthoryear{Xie, Liu, Jia, Luan, and Sun}{Xie
  et~al\mbox{.}}{2016}]%
        {DKRL}
\bibfield{author}{\bibinfo{person}{Ruobing Xie}, \bibinfo{person}{Zhiyuan Liu},
  \bibinfo{person}{Jia Jia}, \bibinfo{person}{Huanbo Luan}, {and}
  \bibinfo{person}{Maosong Sun}.} \bibinfo{year}{2016}\natexlab{}.
\newblock \showarticletitle{Representation Learning of Knowledge Graphs with
  Entity Descriptions}. In \bibinfo{booktitle}{\emph{AAAI}}.
  \bibinfo{publisher}{AAAI Press}, \bibinfo{address}{New York, NY, USA},
  \bibinfo{pages}{2659--2665}.
\newblock


\bibitem[\protect\citeauthoryear{Yang, Yih, He, Gao, and Deng}{Yang
  et~al\mbox{.}}{2015}]%
        {DistMult}
\bibfield{author}{\bibinfo{person}{Bishan Yang}, \bibinfo{person}{Wen{-}tau
  Yih}, \bibinfo{person}{Xiaodong He}, \bibinfo{person}{Jianfeng Gao}, {and}
  \bibinfo{person}{Li Deng}.} \bibinfo{year}{2015}\natexlab{}.
\newblock \showarticletitle{Embedding Entities and Relations for Learning and
  Inference in Knowledge Bases}. In \bibinfo{booktitle}{\emph{ICLR}}.
  \bibinfo{publisher}{OpenReview.net}, \bibinfo{address}{San Diego, CA, USA},
  \bibinfo{pages}{1--12}.
\newblock


\bibitem[\protect\citeauthoryear{Yang, Zhang, Gao, Ji, and Chen}{Yang
  et~al\mbox{.}}{2019}]%
        {RE2}
\bibfield{author}{\bibinfo{person}{Runqi Yang}, \bibinfo{person}{Jianhai
  Zhang}, \bibinfo{person}{Xing Gao}, \bibinfo{person}{Feng Ji}, {and}
  \bibinfo{person}{Haiqing Chen}.} \bibinfo{year}{2019}\natexlab{}.
\newblock \showarticletitle{Simple and Effective Text Matching with Richer
  Alignment Features}. In \bibinfo{booktitle}{\emph{ACL}}.
  \bibinfo{publisher}{ACL}, \bibinfo{address}{Florence, Italy},
  \bibinfo{pages}{4699--4709}.
\newblock


\bibitem[\protect\citeauthoryear{Yin, Han, and Yu}{Yin et~al\mbox{.}}{2008}]%
        {TruthFinder}
\bibfield{author}{\bibinfo{person}{Xiaoxin Yin}, \bibinfo{person}{Jiawei Han},
  {and} \bibinfo{person}{Philip~S. Yu}.} \bibinfo{year}{2008}\natexlab{}.
\newblock \showarticletitle{Truth Discovery with Multiple Conflicting
  Information Providers on the Web}.
\newblock \bibinfo{journal}{\emph{IEEE Transactions on Knowledge and Data
  Engineering}} \bibinfo{volume}{20}, \bibinfo{number}{6}
  (\bibinfo{year}{2008}), \bibinfo{pages}{796--808}.
\newblock


\bibitem[\protect\citeauthoryear{Zhao, Rubinstein, Gemmell, and Han}{Zhao
  et~al\mbox{.}}{2012}]%
        {LTM}
\bibfield{author}{\bibinfo{person}{Bo Zhao}, \bibinfo{person}{Benjamin I.~P.
  Rubinstein}, \bibinfo{person}{Jim Gemmell}, {and} \bibinfo{person}{Jiawei
  Han}.} \bibinfo{year}{2012}\natexlab{}.
\newblock \showarticletitle{A Bayesian Approach to Discovering Truth from
  Conflicting Sources for Data Integration}.
\newblock \bibinfo{journal}{\emph{Proceedings of the VLDB Endowment}}
  \bibinfo{volume}{5}, \bibinfo{number}{6} (\bibinfo{year}{2012}),
  \bibinfo{pages}{550--561}.
\newblock


\bibitem[\protect\citeauthoryear{Zheng, Li, Li, Shan, and Cheng}{Zheng
  et~al\mbox{.}}{2017}]%
        {TISurvery}
\bibfield{author}{\bibinfo{person}{Yudian Zheng}, \bibinfo{person}{Guoliang
  Li}, \bibinfo{person}{Yuanbing Li}, \bibinfo{person}{Caihua Shan}, {and}
  \bibinfo{person}{Reynold Cheng}.} \bibinfo{year}{2017}\natexlab{}.
\newblock \showarticletitle{Truth Inference in Crowdsourcing: Is the Problem
  Solved?}
\newblock \bibinfo{journal}{\emph{Proceedings of the VLDB Endowment}}
  \bibinfo{volume}{10}, \bibinfo{number}{5} (\bibinfo{year}{2017}),
  \bibinfo{pages}{541--552}.
\newblock


\end{thebibliography}

\appendix

\section{Model Complexity}
\label{app:comp}

In fact scoring, the GNN contains $(K-1) d_h^2 + d_h d_e$ parameters, initial entity embeddings contain $\frac{1}{2} N_e d_e$ parameters, attribute embeddings contain $N_{rel} d_e^2 + N_{lit} (d_e + 1)$ parameters, and string encoder contains $d_{BERT} d_e$ parameters, where $d_h$ is the dimension of GNN hidden layers, $N_e$ is the number of entities, $N_{rel}$ is the number of attributes used in relational fact scoring, $N_{lit}$ is the number of attribute used in literal fact scoring, and $d_{BERT}$ is the output dimension of BERT. 

In value alignment networks, the literal-literal alignment network contains $8 d_{BERT}^2 + d_{BERT} d_{ll} + d_{ll}$ parameters, where $d_{ll}$ is the dimension of its hidden layer. 
The literal-entity alignment network contains $(d_{BERT}+d_e) d_{le}$ parameters, where $d_{le}$ is the dimension of its hidden layer. 

The truth inference contains $N_{lit} + N_{rel} + N_s + d_e d_{de}$ parameters, where $N_s$ is the number of data sources, and $d_{de}$ is the hidden layer dimension of entity difference function.

In total, the parameter number of our method is $(K-1) d_h^2 + d_h d_e + d_e(\frac{1}{4}d_e+\frac{1}{2}N_e+N_{rel}+d_{BERT}+N_{lit}) + N_{lit} + d_{ll} (d_{BERT}+ 1) + 8 d_{BERT}^2 + d_{le}(d_{BERT}+d_e) + N_{lit} + N_{rel} + N_s + d_e d_{de}$.

\section{Hyperparameter Setting}
\label{app:hp}

We search the following hyperparameter values for model training: the learning rate in $\{0.0001, 0.0005, 0.001, 0.005, 0.01\}$ and the batch size in $\{64, 128, 256, 512, 1024\}$. 
The selected learning rate is $0.005$, and the picked batch size is $128$.

For holistic fact scoring, we try the number $K$ of GNN layers in $\{1, 2, 3, 4\}$, the dimension of hidden layers in $\{100, 200, 300, 400, 500\}$, and the dimensions of entity and value embeddings in $\{25, 50, 100,$ $200, 300\}$. 
We randomly sample $50$ neighbors for each entity, and choose a two-layer GNN (i.e., $K=2$) with the dimension of  hidden layers $100$.
The dimensions of entity and value embeddings are both set to $100$.
The dimension of relational attribute embeddings is $100 \times 100$, and the dimension of literal attribute embeddings is $100$. 
The threshold for value set size is $N_v = 10$.


For value alignment networks, we attempt the dimension of MLP hidden layers in $\{100, 200, 300, 400, 500\}$. The chosen dimension for both literal-literal alignment network and literal-entity alignment network is $100$.

For semi-supervised truth inference, we test the output dimension of linear transform $\mathbf{W}_e$ in the difference function in $\{25, 50, 75,$ $100\}$, and set $\mathbf{W}_e = 100 \times 25$. The $FACT\_EPOCHS$ is set to $20$, and $INFERENCE\_EPOCHS$ is also set to $20$.

\section{Runtime Analysis}
\label{app:rt}

We compare the overall runtime of TKGC with two existing open KG completion methods, OKELE and Knowledge Vault, as well as two variants of TKGC, namely TKGC w/o $\mathcal{F}_{lit}$ and TKGC w/o $\mathcal{F}_{rel}$.

As shown in Table~\ref{tab:exp_runtime}, we have the following findings: (i) OKEKE runs relatively faster because it only uses facts in the KG for property prediction, but does not use them for truth inference. Differently, TKGC learns prior knowledge inside the KG with fact scoring to enhance truth inference, making it spend more time. (ii) Unlike Knowledge Vault that only considers the trustworthiness of facts, TKGC additionally considers the trustworthiness of data sources. The more complex truth inference model causes TKGC to run a little slower than Knowledge Vault. (iii) Compared with TKGC w/o $\mathcal{F}_{lit}$, TKGC spends 0.3 minutes to process 5,727 literal facts. Compared with TKGC w/o $\mathcal{F}_{rel}$, TKGC spends 4.8 minutes to process 186,032 relational facts. The remaining runtime of TKGC is mainly spent by the GNN module that aggregates the neighboring information of entities.

\begin{table}[!ht]
\centering
\caption{Runtime of open KG completion methods}
\label{tab:exp_runtime}
{\small
    \begin{tabular}{l|cc|ccc}
        \toprule    Methods & Runtime (min.) \\
        \midrule    OKELE & \ \ 7.3 \\
                    Knowledge Vault & 16.5 \\
        \midrule    TKGC w/o $\mathcal{F}_{lit}$ & 17.8 \\
                    TKGC w/o $\mathcal{F}_{rel}$ & 13.3 \\
                    TKGC (full) & 18.1 \\
        \bottomrule
    \end{tabular}}
\end{table}

\end{document}